\documentclass[aps,prb,twocolumn,superscriptaddress]{revtex4-2}
\usepackage{tabularx} 
\usepackage{amsmath}  
\usepackage{graphicx} 
\usepackage[margin=1in,letterpaper]{geometry} 
\usepackage[final]{hyperref} 
\usepackage[normalem]{ulem}
\usepackage{siunitx}
\hypersetup{
	colorlinks=true,       
	linkcolor=blue,        
	citecolor=blue,        
	filecolor=magenta,     
	urlcolor=blue         
}
\usepackage{blindtext}

\begin{document}

\title{Breakdown of the drift-diffusion model for transverse spin transport in a disordered Pt film}
\author{K. D. Belashchenko}
\author{G. G. Baez Flores}
\author{W. Fang}
\author{A.~A.~Kovalev}

\affiliation{Department of Physics and Astronomy and Nebraska Center for Materials and Nanoscience, University of Nebraska-Lincoln, Lincoln, Nebraska 68588, USA}

\author{M. van Schilfgaarde}
\affiliation{National Renewable Energy Laboratory, Golden, Colorado 80401, USA}

\author{P. M. Haney}
\author{M. D. Stiles}

\affiliation{Physical Measurement Laboratory, National Institute of Standards and Technology, Gaithersburg, Maryland 20899, USA}

\date{\today}

\begin{abstract}
Spin accumulation and spin current profiles are calculated for a disordered Pt film subjected to an in-plane electric current within the nonequilibrium Green function approach. In the bulklike region of the sample, this approach captures the intrinsic spin Hall effect found in other calculations. Near the surfaces, the results reveal qualitative differences with the results of the widely used spin-diffusion model, even when the boundary conditions are modified to try to account for them. One difference is that the effective spin-diffusion length for transverse spin transport is significantly different from its longitudinal counterpart and is instead similar to the mean-free path. This feature may be generic for spin currents generated via the intrinsic spin-Hall mechanism because of the differences in transport mechanisms compared to longitudinal spin transport. Orbital accumulation in the Pt film is only significant in the immediate vicinity of the surfaces and has a small component penetrating into the bulk only in the presence of spin-orbit coupling, as a secondary effect induced by the spin accumulation.
\end{abstract}

\maketitle

\section{Introduction}

In non-magnetic materials, the spin Hall effect \cite{sinova2015spin} describes the spin current flowing perpendicular to an applied electric field. Originally predicted in 1971 \cite{dyakonov1971current}, this effect attracted a lot of interest around the beginning of this century because of three developments. It was indirectly observed in semiconductors through the resulting spin accumulation at the edges of the sample \cite{kato2004observation}. The theoretical community described a new ``intrinsic'' contribution that was independent of the scattering mechanisms and could be calculated just from the properties of the bulk electronic structure \cite{murakami2004spin} building on related developments in the anomalous Hall effect in ferromagnets \cite{nagaosa2010anomalous}. The final development was the proposal that the spin Hall effect plays a crucial role in spin-orbit torques \cite{liu2011spin}. Spin-orbit torques are a feature of bilayers of a ferromagnet and a nonmagnetic layer, typically a heavy metal with strong spin-orbit coupling. An electric field in the plane of the sample gives rise to a spin Hall current perpendicular to the sample plane toward the interface. That spin current then exerts a torque on the magnetization of the ferromagnet. The possibility of applications of spin-orbit torques \cite{shao2021roadmap} continues to drive both experimental and theoretical research.

The intrinsic contribution to the spin Hall effect describes the transverse spin current for cases of moderate disorder scattering. Detailed transport calculations show that in the clean limit, the spin Hall conductivity increases above the intrinsic value due to increasing transport time between scattering events \cite{sinova2015spin} so it is instead described by extrinsic scattering, like skew scattering. This counterintuitive situation makes the intrinsic spin Hall conductivity interesting in that it is a property of the band structure of the disorder-free material but manifests itself only when the disorder scattering is strong enough that other conduction mechanisms do not dominate. 

Relating measured spin-orbit torques \cite{manchon2019current} to the spin Hall effect requires a model for spin transport. In most cases, the analysis uses a drift-diffusion approach as done initially in Ref.~\cite{liu2011spin} and described in Sec~\ref{sec:SpinDiffusion}. Issues associated with this approach include the following. Does such an approach include the appropriate length scales? Since the discovery of giant magnetoresistance, it has been known \cite{camley1989theory} that the mean-free path can be the dominant length scale for in-plane spin transport. This length scale is not captured in a drift-diffusion approach, where the spin-diffusion length is the only length scale. Can an intrinsic spin Hall current be simply added as a source term, or do corrections need to be included? Does the neglect of the details of the electronic structure matter?  The Fermi surface can be complicated, and attendant quantities such as the velocity and Berry curvature can vary widely at different points on the surface.  Finally, what are the appropriate boundary conditions? Typically, researchers assume that the boundary conditions for the spin transport are described by the spin-mixing conductance from magnetoelectronic circuit theory \cite{brataas2006non}. Some additional contributions to the boundary conditions have been recognized by model calculations using the Boltzmann equation \cite{haney2013currenta} and the subsequent analysis that led to the introduction of additional terms \cite{amin2016formalism,amin2016phenomenology}. One of these additional terms was incorporated in the semiclassical analysis of a first-principles calculation closely related to the present one \cite{nair2021spin}. However, the justification for describing the intrinsic spin-Hall current near the surface using such additional boundary terms is lacking. As we discuss below, the assumptions of the spin diffusion model may be too limited to allow for boundary conditions to fully describe the details of spin transport at the interface without introducing additional degrees of freedom.

First-principles calculations of spin-orbit torques in transition-metal bilayers \cite{haney2013currentb,freimuth2014spin,belashchenko2019first,mahfouzi2020microscopic,flores2022effect,fang2022first,go2020theory} do not make the assumptions used in drift-diffusion approaches, but such calculations are generally difficult to use to analyze experiments. They are computationally quite intensive, making it difficult to perform systematic studies. The calculations require as input a specific model for the structure of the samples, but disorder is difficult to include. In most experimental measurements of spin-orbit torques, the disorder is strong enough that the measured resistivities are much larger than those typically measured in bulk samples of the same materials. Detailed measurements of this disorder are lacking. It is also prohibitive to include the materials typically capping both sides of the bilayer structures. 

In this paper, we build on an earlier calculation \cite{nair2021spin} to address the uncertainty in spin transport in heavy-metal layers. To focus on the spin transport due to the spin Hall effect, we leave out the ferromagnetic layer and consider an isolated slab of Pt with two vacuum surfaces. The calculation uses the non-equilibrium Green function (NEGF) formalism \cite{Datta1997,Nikolic2018} with onsite Anderson disorder implemented within the linear muffin-tin orbital (LMTO) method in the Questaal code \cite{Faleev2005,belashchenko2019first,pashov2020questaal}. While this approach is distinct from the wavefunction-based scattering formalism used by Nair \emph{et al.}~\cite{nair2021spin}, it is fundamentally equivalent and should capture the same physics. At present, only approaches like these, with Landauer-style embedding and explicit averaging over disorder, are feasible for calculating diffusive quantum transport in systems with surfaces or interfaces. We treat the case with strong enough disorder that the intrinsic mechanism is expected to dominate over extrinsic mechanisms. 

Our main results can be summarized as follows. The NEGF approach captures the intrinsic spin Hall effect in the bulklike region of the sample. This conclusion is based on the agreement between the calculated magnitude of the spin current and the calculations of the bulk intrinsic contribution in the literature \cite{Guo2008,tanaka2008intrinsic,salemi2022first} and the insensitivity of the spin-Hall conductivity to the strength of the disorder. Similar to spin-orbit torques in a Co/Pt bilayer \cite{belashchenko2019first}, the spin-Hall current in Pt is dominated by the Fermi surface contribution with only a minor contribution from the Fermi sea. Near the surfaces, our calculations are inconsistent with semiclassical models of transport. First, the exponential decay lengths for deviations from bulk behavior near the surfaces (transverse spin-diffusion lengths) are much shorter than the longitudinal spin-diffusion length and are close to the mean free path. The behavior of the bulk intrinsic spin current near a surface requires a modification of the boundary conditions. The surface absorbs part of the intrinsic spin current directly into the lattice, while also generating extrinsic spin current and accumulation. The spin-accumulation-free absorption needs to be included in the boundary condition. Finally, even with the updated boundary conditions, the decay of the surface spin accumulation and of the excess spin current near the surface are not consistent with the assumptions of the semiclassical model.

Section~\ref{sec:SpinDiffusion} presents the standard drift-diffusion model used to analyze spin-orbit torque measurements. We extend the boundary conditions used in the model in a way that is essential to analyze the calculations presented below. Section~\ref{sec:computational} describes the computational details, including the implicit calculation of the spin current from the spin continuity equation by spatially integrating its divergence. Section~\ref{sec:transverse} gives the results of the calculations of the transverse transport. First, Sec.~\ref{sec:accumulation} gives the spin accumulation, which can be computed directly, and then Sec.~\ref{sec:current} gives the spin current. The results of both sections are fit to the drift-diffusion calculation described in Sec.~\ref{sec:SpinDiffusion}. The comparison of the fits shows that the drift-diffusion model is not consistent with the numerical results. Sec.~\ref{sec:discuss} discusses the calculations and results in the context of other related calculations and the implications of the results for the interpretation of experiments. Sec.~\ref{sec:orbital} describes the calculated orbital accumulations and what we can conclude from it. Finally, we summarize the paper in Sec.~\ref{sec:conclusions}.

\section{Spin diffusion model}
\label{sec:SpinDiffusion}

We consider a nonmagnetic film of thickness $d$ filling the space at $-d/2<z<d/2$ with a homogeneous charge current density $j=\sigma E$ flowing along the $x$ axis.
The spin current flows in the $z$ direction and is polarized along the $y$ axis. Its density is
\begin{equation}
    j_s=j_\mathrm{SH}+j_\mathrm{BF}=\frac{\hbar}{2e}\left(\sigma_\mathrm{SH} E -\sigma_s\frac{\partial \mu_y}{\partial z}\right)
    \label{js}
\end{equation}
where the first term is the bulk spin-Hall current and the second is the ``spin backflow'' associated with spin diffusion. Here $\sigma_s=\sigma/2$ is the conductivity per spin, and $\mu_y$ is the $y$-component of the spin accumulation \cite{Zutic2004RMP}.

The steady-state spin continuity equation is
\begin{equation}
    dj_s/dz = -\frac{\hbar}{2} \frac{n_s}{\tau_\mathrm{sf}}
    \label{continuity}
\end{equation}
where $\hbar n_s/2$ is the spin density (polarized along $y$) per unit volume, and $\tau_\mathrm{sf}$ the spin relaxation time. 
Substituting (\ref{js}) into (\ref{continuity}), assuming $\sigma_\mathrm{SH}$ is homogeneous, and using the relation $N_se\mu_y=n_s$ (where $N_s$ is the density of states per spin per unit volume), we find
\begin{equation}
    \frac{d^2 \mu_y}{dz^2}= \frac{\mu_y}{l_\mathrm{sf}^2}
    \label{sde}
\end{equation}
where $l_\mathrm{sf}=(D\tau_\mathrm{sf})^{1/2}$ is the spin-diffusion length and $D=\sigma_s/(N_se^2)$.
In the following we will use the notation $\bar{\jmath}_s=2ej_s/\hbar$.

For the discussion below, it is useful to call attention to two major assumptions of this approach, at least one of which must fail in the first-principles results we present below. The second term in parentheses in Eq.~\ref{js} assumes that the details of the nonequilibrium distribution function do not matter, or, in other words, that all electrons contribute similarly to the conductivity. For d-band metals, properties like the velocity, Berry curvature, and scattering rates vary widely over the Fermi surface. The right-hand side of Eq.~\ref{continuity} assumes that the relaxation of spin to orbital moments and then to the lattice proceeds only through the spin expectation value $\langle n_s\rangle$ and not through processes that depend on quantities like $\langle \mathbf{s} \times \mathbf{l} \rangle$, where $\mathbf{l}$ is the orbital angular momentum operator.

The boundary condition at the surface of the Pt film needs to account for surface spin relaxation as well as the possible generation or absorption of spin current. The proper boundary condition within the spin diffusion model cannot be derived without explicit knowledge of scattering at the interface; and as we show later, the NEGF theory 
indicates that the semiclassical transport equations are violated. In any case, it is unclear whether the spin diffusion model can describe the intrinsic spin current in the first place.  We try the phenomenological boundary condition
\begin{equation}
    \bar{\jmath}_s|_{z=d/2} = G_{sl} \mu|_{z=d/2} + \sigma_\mathrm{SSH} E
    \label{bc}
\end{equation}
where $G_{sl}$ is the surface spin-loss conductance and $\sigma_\mathrm{SSH}$ the ``surface spin-Hall conductivity'' which gives the spin current injected into the bulk from the surface. The first term in the boundary condition is similar to the spin memory loss term that has been added~\cite{SML2016} to the original interface conductance matrix in magnetoelectronic circuit theory~\cite{brataas2006non}. Here, it describes the difference between the incoming and outgoing spin current that arises from the relaxation of the spin accumulation at the surface. The second term describes the spin current generated or absorbed at the interface due to the electric field, similarly to the generation of a spin Hall current in the bulk. If $\sigma_\mathrm{SSH}$ has the same sign as $\sigma_\mathrm{SH}$, it means that part of the bulk spin-Hall current is absorbed by the surface rather than being compensated by diffusive backflow from the spin accumulation.

The solution of the spin diffusion equation (\ref{sde}) with the boundary condition (\ref{bc}) is
\begin{equation}
    \frac{\mu_y(z)}{E}=\frac{2(\theta_\mathrm{SH}-\theta_\mathrm{SSH})l_\mathrm{sf}\sinh{z/l_\mathrm{sf}}}{\cosh d/(2l_\mathrm{sf}) + g_{sl} \sinh d/(2l_\mathrm{sf})}
    \label{muy}
\end{equation}
where $\theta_\mathrm{SH}=\sigma_\mathrm{SH}/\sigma$ and $\theta_\mathrm{SSH}=\sigma_\mathrm{SSH}/\sigma$ are the bulk and surface spin-Hall angles, and $g_{sl}=2G_{sl}l_\mathrm{sf}\rho$ is the dimensionless spin-loss parameter. We rewrite (\ref{muy}) as
\begin{equation}
    \frac{\mu_y(z)}{E}=2l_\mathrm{sf}\theta_\mathrm{eff}\frac{\sinh{z/l_\mathrm{sf}}}{\cosh d/(2l_\mathrm{sf})}
    \label{muy2}
\end{equation}
where
\begin{equation}
    \theta_\mathrm{eff}=\frac{(\theta_\mathrm{SH}-\theta_\mathrm{SSH})}{1 + g_{sl} \tanh d/(2l_\mathrm{sf})}
    \label{theff}
\end{equation}
can be called the effective spin-Hall angle for spin accumulation. $\theta_\mathrm{eff}$ is equal to $\theta_\mathrm{SH}$ if the surface does not absorb or emit any spin current, i.e., if the right-hand side in Eq.~(\ref{bc}) vanishes. The spin current is
\begin{equation}
    \frac{\bar{\jmath}_s(z)}{j}=\theta_\mathrm{SH}-\theta_\mathrm{eff}\frac{\cosh{z/l_\mathrm{sf}}}{\cosh d/(2l_\mathrm{sf})} .
    \label{jssol}
\end{equation}
Note that in the case $\theta_\mathrm{SSH}=\theta_\mathrm{SH}$ (regardless of $g_{sl}$) the effective spin-Hall angle vanishes, the spin accumulation is zero everywhere, while the spin current is constant across the whole thickness of the film, ``disappearing'' at the surface.

Deep in the bulk of a thick film ($d\gg l_\mathrm{sf}$), the spin current ratio $\bar{\jmath}_s/j$  is equal to $\theta_\mathrm{SH}$, but the spin accumulation near the surface is instead proportional to $\theta_\mathrm{eff}$. The second term in (\ref{jssol}) comes from the spin backflow and depends on the same $\theta_\mathrm{eff}$. 
Whether the spatial profile of the spin accumulation (\ref{muy2}) and spin current (\ref{jssol}) are described by the same $\theta_\mathrm{eff}$ can be used to test the validity of the spin diffusion model.

Although the two boundary terms that depend on $\sigma_\mathrm{SSH}$ and $G_{sl}$ respectively affect the spin accumulation differently, it is impossible to determine exactly the role each plays. It is possible in some cases to determine that one of them cannot explain the measured behavior alone. In the limit $d\gg l_\mathrm{sf}$ we have $\mu_y(d/2)/E=2l_\mathrm{sf}\theta_\mathrm{eff}$, where $\theta_\mathrm{eff}\approx(\theta_\mathrm{SH}-\theta_\mathrm{SSH})(1+g_{sl})^{-1}$. With increasing disorder, $\rho$ increases while $l_\mathrm{sf}$ decreases. If the spin-flip lifetime $\tau_\mathrm{sf}$ is proportional to the momentum relaxation time $\tau$, which is expected for both impurity and phonon scattering \cite{Elliott} and should hold in our Anderson disorder model, then $l_\mathrm{sf}$ is proportional to $\tau$. If the bulk spin-Hall effect is intrinsic (that is, $\sigma_\mathrm{SH}$ is independent of disorder), then $\theta_\mathrm{SH} l_\mathrm{sf}\sim\mathrm{const}$, and the dependence of $\mu_y(z/2)/E$ on disorder depends on $(1+g_{sl})^{-1}$ and the surface spin-Hall angle $\theta_\mathrm{SSH}$. If the spin-Hall effect comes from skew scattering on disorder (both in the bulk and at the surface), then $\theta_\mathrm{eff}\sim\mathrm{const}$, and $\mu_y(d/2)/E\propto\tau$.
On the other hand, for a thin film ($d \ll l_\mathrm{sf}$) with small spin memory loss ($g_{sl}\le 1$) the spin accumulation at the surface $\mu_y(d/2)/E\approx (\theta_\mathrm{SH}-\theta_\mathrm{SSH})d$ does not depend on $g_{sl}$.

\section{Computational approach}
\label{sec:computational}

The Pt film is oriented so that its free surfaces are normal to the [001] direction, the bias is applied across the [110] direction, and the sample is periodic in the remaining [1$\overline{1}$0] direction of the face-centered cubic (fcc) lattice.
In this geometry, it is more convenient to view the fcc lattice as an equivalent body-centered tetragonal (bct) lattice in which the $c$ axis along [001] is unchanged and the fcc [110] and [1$\overline{1}$0] directions become bct [100] and [010], respectively, with $c/a=\sqrt{2}$.
We consider the active (scattering) region with dimensions of $100\times300\times2$ monolayers (ML) in these three directions ($50\times150\times1$ bct unit cells or $19.4\times41.6\times0.52$ nm). The small width of the film in the third direction is restricted by computational limitations. In the [001] direction, periodically repeated copies of the film are separated by 4 ML of empty spheres, which represent vacuum.

Disorder is treated within the Anderson model by applying a random potential to each of the 15600 distinct atomic sites explicitly treated in the system. At each site, the disorder potential $V_i$ is randomly chosen from a uniform distribution bounded by $|V_i|<V_m$. All calculations in the paper are averaged over multiple independent realizations of this disorder potential.

The active region is embedded between semi-infinite leads which represent the continuation of the same Pt film without disorder.
Our goal is to study diffusive transport, which sets in within a couple of mean-free paths away from the embedding planes where the disordered active region is connected to ideal leads. Therefore, physical quantities, such as spin and orbital accumulations and spin-orbital torques, are accumulated only over the 180 ML in the center of the active region. This excludes 60 ML (8.3~nm) adjacent to each lead. Thus, the results reflect diffusive transport properties of the film in the Ohmic limit as long as the mean-free path $\lambda$ does not exceed a few nanometers.

Using the NEGF approach \cite{Datta1997,Nikolic2018,Faleev2005,belashchenko2019first,pashov2020questaal},
we compute the linear responses to the bias voltage drop $V$ between the left and right reservoirs:
$\mathbf{t}_i = e^{-1} d\mathbf{T}_i/dV$, where $\mathbf{T}_i$ is the spin-orbital torque on site $i$ or another local observable such as the spin accumulation.
We need to convert these results to the linear response $\boldsymbol{\tau}_i=d\mathbf{T}_i/dE$ to the electric field $E$ inside an extended sample, such as the torquance.
To find the electric field in the bulk-like central region of the device, we include a correction for the potential drop in the contacts. This is done by writing $E=V/L_\mathrm{eff}$ with $L_\mathrm{eff}=R(dR/dL)^{-1}$ where $R=1/G_\mathrm{LB}$ is the Landauer-B\"uttiker resistance and $L$ the length of the active region. We then obtain 
$\boldsymbol{\tau}_i = e L_\mathrm{eff} \mathbf{t}_i$, 
which applies to any local observable. The derivative $dR/dL$ is obtained from the calculations of the Landauer-B\"uttiker conductance $G_\mathrm{LB}$ as a function of $L$.

\begin{figure}
    \centering
    \includegraphics[width=0.85\columnwidth]{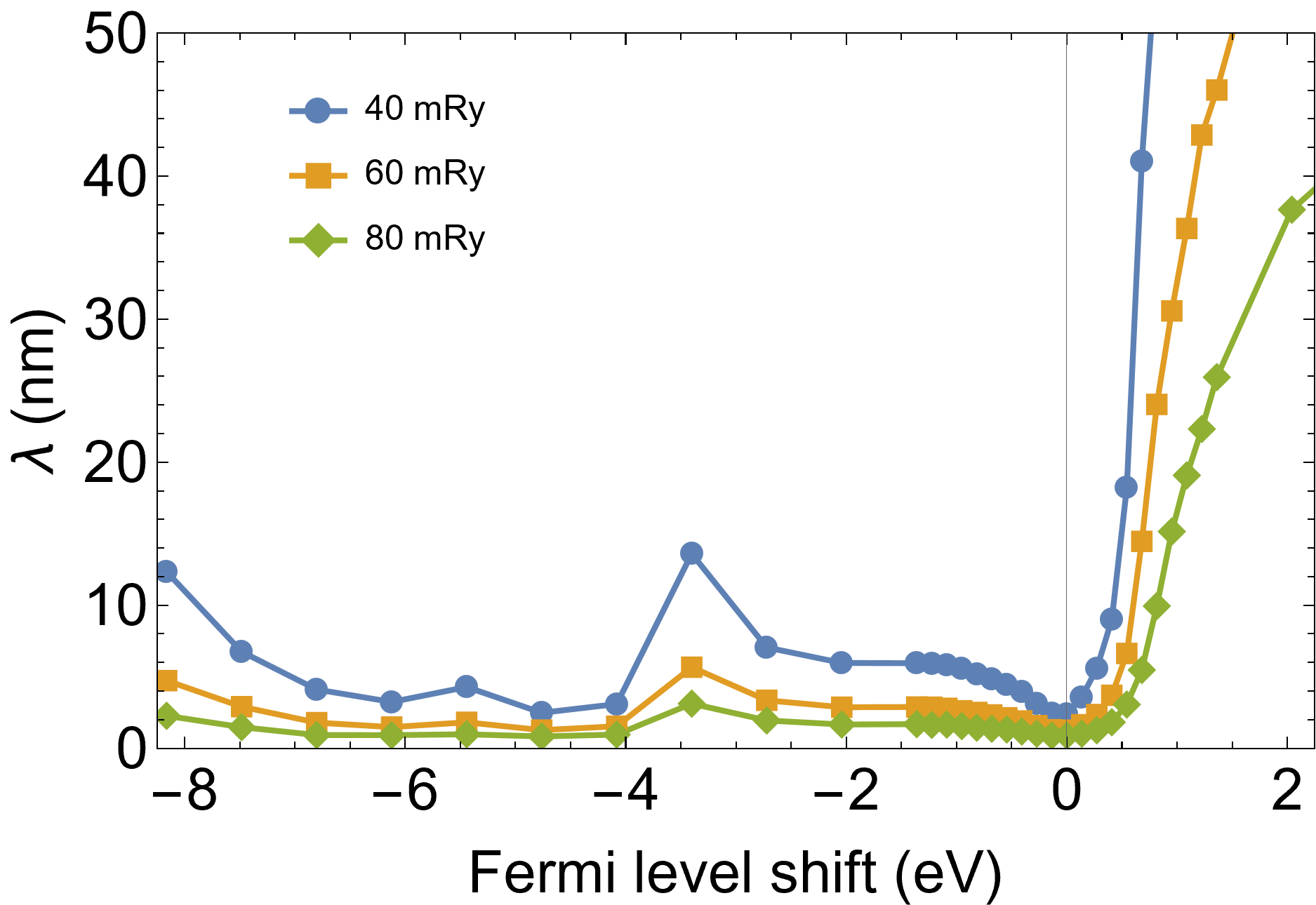}
    \caption{Mean-free path $\lambda$ in Pt, as a function of the hypothetical Fermi level shift relative to the true Fermi level, at different disorder strengths.}
    \label{fig:mfp}
\end{figure}

To investigate the generality of our results, we consider model materials accessed by artificially varying the Fermi level and by varying the disorder. Varying the Fermi level illustrates how different features of the band structure affect the results. To determine the role of disorder, we vary $V_m$, the bounds on the values of the disorder potentials, through the values 0.544~eV, 0.816~eV, 1.088~eV, and 1.361~eV, which are 40~mRy, 60~mRy, 80~mRy, and 100~mRy respectively, where the Rydberg unit of energy is 1~Ry=13.606~eV. For clarity, in the rest of the paper, we refer to these values of disorder in these atomic units. Figure~\ref{fig:mfp} shows the energy and disorder dependence of $\lambda$ obtained using separate calculations of the conductivity and the mean-squared Fermi velocities, assuming constant relaxation time \cite{Gall2016,Nair2021}. From Table~\ref{tab:coefs}, we see that $\lambda$ is consistent with the requirement of being less than 8.3~nm as long as the Fermi energy does not go more than about 1 eV above the true Fermi level. This is where the $5d$ character of the bands fades away, and the scattering cross-section is suppressed by the large bandwidth of the $s$ electrons. The mean free path $\lambda$ is also fairly large at $E_F\approx-3.5$ eV, especially at $V_m=40$ mRy. As expected, $\lambda$ decreases with increasing disorder strength, i.e.~increasing $V_m$. The short mean free paths are consistent with the calculations being done in the large disorder limit, in which the intrinsic spin Hall effect is expected to dominate over the extrinsic spin Hall effect.

In the steady state, the divergence of the spin current is balanced by the total torque acting on the spin, which includes contributions from the exchange torque and the spin-orbital torque \cite{HaneyStiles2010,go2020theory}. Note that the term \emph{spin-orbital torque} refers to the torque on the spin density from the orbital moments through spin-orbit coupling. This usage is in contrast to the term \emph{spin-orbit torque} that refers to the torque on the magnetization. In density functional theory, these torques are obtained by commuting the spin density operator with operators representing the spin-orbit coupling and the exchange-correlation field, respectively. The exchange torque vanishes in a nonmagnetic metal, and hence the divergence of the spin current is balanced by the spin-orbital torque alone. Within the rigid-spin approximation \cite{RSA} the spin-orbital torque $\mathbf{T}_i^{SO}$ is a site-resolved quantity, and we can work with a discretized spin current $\mathbf{I}_s$ such that
\begin{equation}
    \mathbf{T}_i^{SO}+\nabla_i \mathbf{I}_s =0
    \label{eq:dsdt}
\end{equation}
where the discrete divergence $\nabla_i\mathbf{I}_s$ gives the influx of spin current into site $i$. Here the spin current flows in the $z$ direction, perpendicular to the film plane, and its vector character refers to its spin polarization. If the spin current flowing between layers 0 and 1 is known to be zero (for example, if 0 is in vacuum), then we can find the spin current flowing between layers $i$ and $i+1$ by summing over the torques
\begin{equation}
    \mathbf{I}_{i,i+1}=\sum_{j\leq i}\mathbf{T}_j^{SO} .
    \label{eq:spincurrentdifference}
\end{equation}
Deep in the bulk (i.e., a few spin-diffusion lengths away from any interfaces) the spin current should correspond to the bulk spin-Hall conductivity $\sigma_\mathrm{SH}$ of the material.

The total spin-orbital torque acting on the spin density of the entire film deviates from zero for a particular disorder sample of a finite supercell, reflecting a small imbalance between the spin current flowing into the active region from one lead and out into the other. In other words, the net steady state transfer of angular momentum between the lattice and the spin current does not cancel as it would for complete disorder averaging. The small residual total torque is averaged over the volume and subtracted from $t^{SO}_{jy}$. This ensures that the spin current $\mathbf{I}_{i,i+1}$ has the periodicity of the supercell.

In the NEGF formalism, the local quantities generally include Fermi-surface and Fermi-sea contributions \cite{belashchenko2019first}. In the case of a Pt film, the Fermi-sea contribution to the spin density is forbidden by time-reversal symmetry, but it is allowed for the spin current. Our estimates described in Appendix~\ref{Fermi-sea} suggest that this contribution amounts to less than 10~\% of the Fermi-surface contribution. In the following, therefore, we neglect the Fermi-sea term.

\section{Transverse spin diffusion}
\label{sec:transverse}

\subsection{Spin accumulation}
\label{sec:accumulation}

In the rest of the paper, we calculate transport properties of Pt slabs at two energies, the nominal Fermi energy $E_F$ and an artificial Fermi energy shifted to $E'=E_F-$2.72~eV (0.2~Ry). Calculations at this second energy allow some insight into the generality of the results.

\begin{figure}[hbt]
    \centering
    \includegraphics[width=0.85\columnwidth]{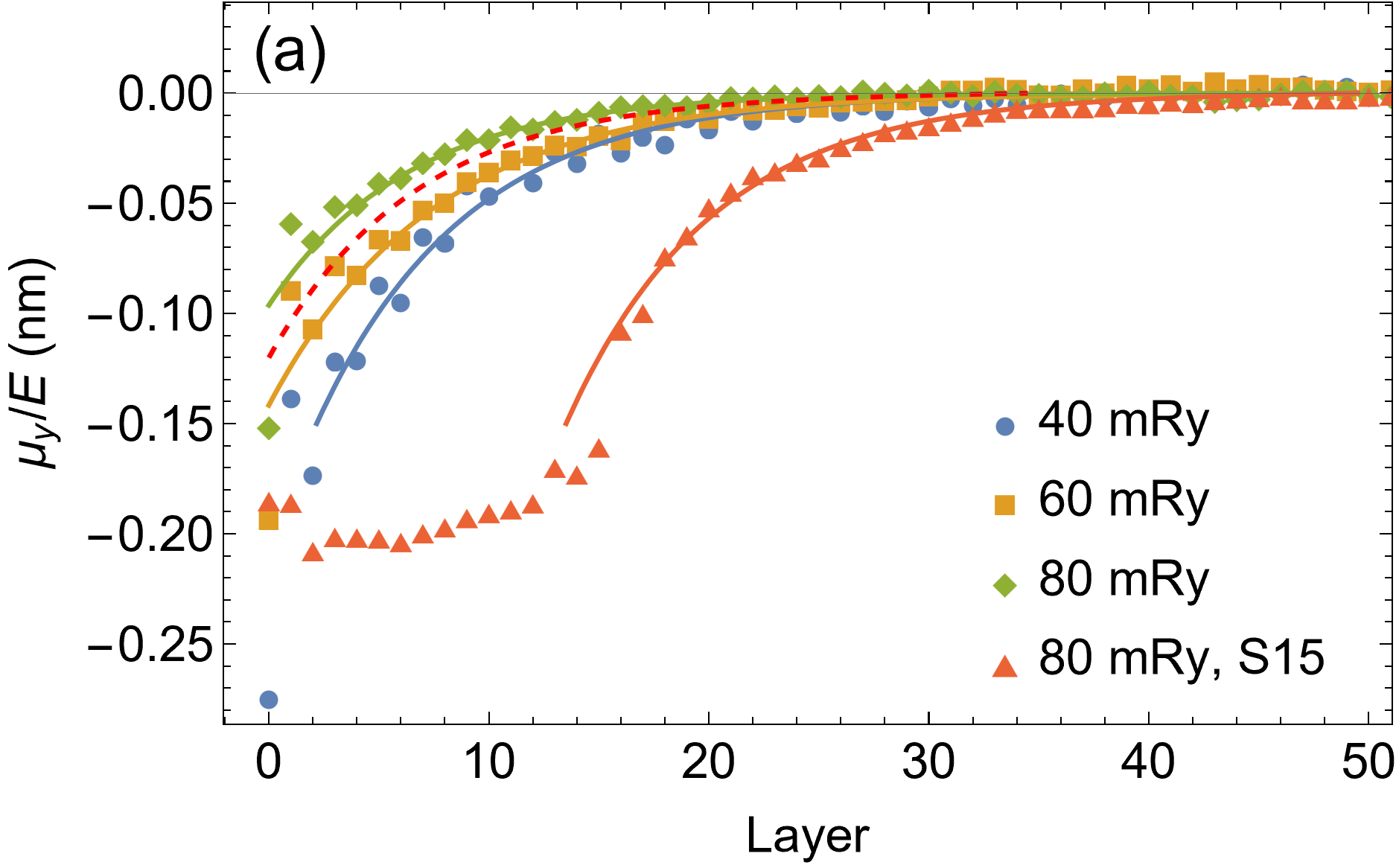}
    \vskip1ex
    \includegraphics[width=0.87\columnwidth]{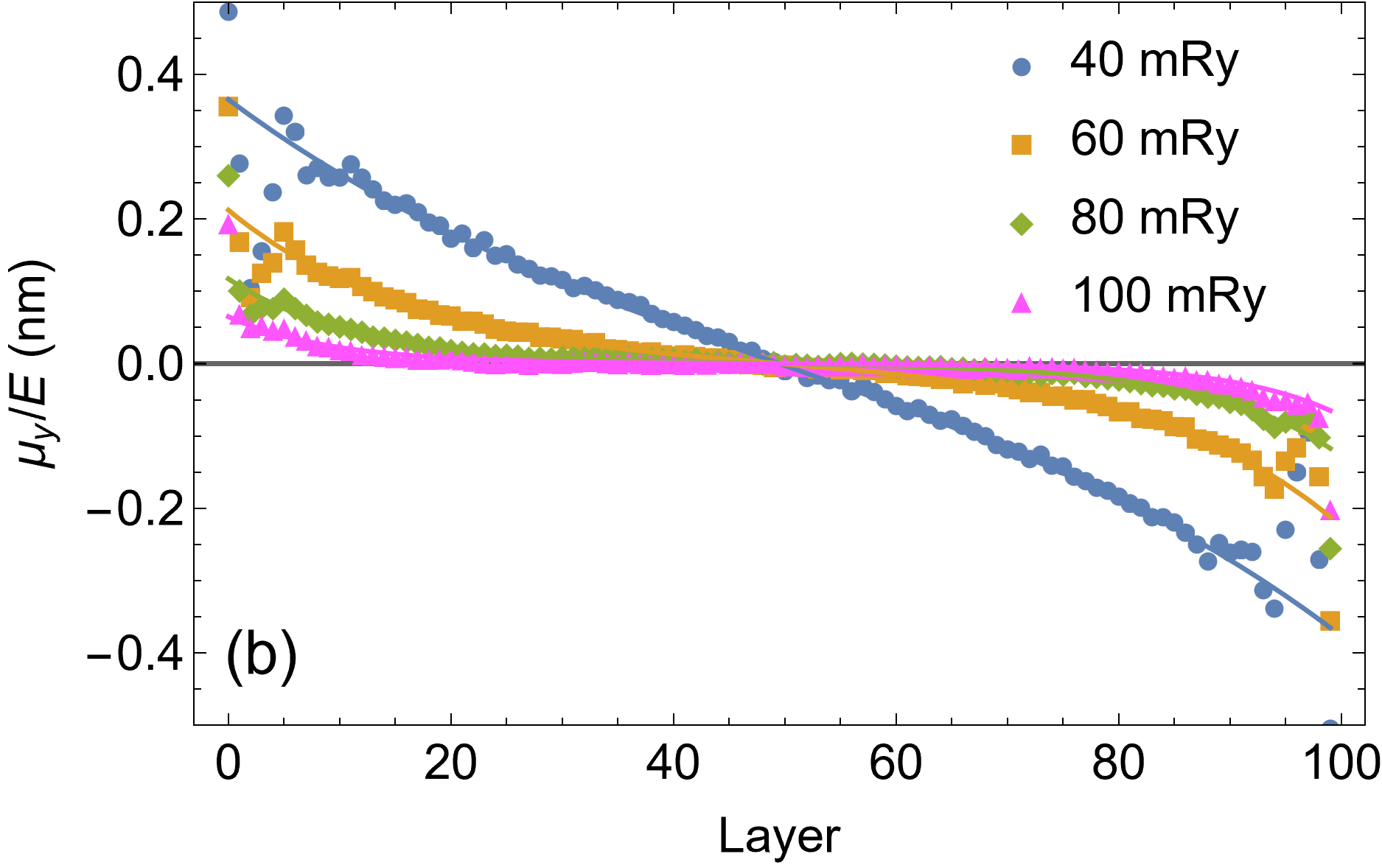}
    \caption{Spin accumulation profile $\mu_y(z)/E$ in a 100-ML film of Pt with Anderson disorder strengths shown in the legends, calculated at (a) the physical Fermi level, or (b) with the Fermi level shifted to \SI{-2.72}{\electronvolt}. Only part of the film is shown in panel (a). Solid lines are fits to Eq.~(\ref{muy2}). The red line marked S15 in panel (a) corresponds to the calculation with spin-orbit coupling turned off in 15 ML of Pt near each interface. The dashed red line is shifted to the left by 15 ML.}
    \label{fig:pt100sy}
\end{figure}

Figures~\ref{fig:pt100sy}(a) and (b) show the spin accumulation profile $\mu_y(z)/E$ in the Pt film calculated for several disorder strengths. Panel (a) corresponds to the physical Fermi level, and panel (b) to the artificially shifted Fermi level $E'$. The data was fit to Eq.~(\ref{muy2}) with the parameters $\theta^{(\mu)}_\mathrm{eff}$ and $l^T_\mathrm{sf}$ listed in Table~\ref{tab:coefs}, which also includes bulk resistivities calculated separately using $16\times2$ ML supercells where the 2-ML width matches the supercells used in the film calculations. Details of the fitting procedure are given in Appendix~\ref{sec:Fitting}.

At $E=E_F$ the fitted transverse spin-diffusion length $l^T_\mathrm{sf}$ is essentially independent of the resistivity, but at $E=E'$ it is roughly proportional to the conductivity. As mentioned above, $l^T_\mathrm{sf}\rho$ is approximately constant as is expected for the Elliott-Yafet mechanism. In contrast, $l^T_\mathrm{sf}$ being approxiately constant at $E=E_F$ is unusual. This behavior of the spin-diffusion length is typical of the Dyakonov-Perel spin relaxation mechanism, which could be dominant near the surface where the inversion symmetry is broken.

Further, we see that $l^T_\mathrm{sf}$ differs from the longitudinal spin-diffusion length $l^L_\mathrm{sf}$ characterizing the decay length of spin current created by mechanisms unrelated to the spin-Hall effect, as described in  Appendix~\ref{sec:logitudinal}.  At $E=E_F$, the difference is more than a factor of 5 at $V_m=40$ mRy, but it decreases with increasing $V_m$ as $l^L_\mathrm{sf}$ declines and approaches the nearly constant value of $l^T_\mathrm{sf}$. At $E=E'$ both $l^L_\mathrm{sf}$ and $l^T_\mathrm{sf}$ decline with increasing disorder, while the $l^L_\mathrm{sf}/l^T_\mathrm{sf}$ ratio declines from about 4 to 2.4 as $\rho$ is increased from \qtyrange{4.5}{28}{\micro\ohm\centi\meter}.

The large difference between $l^L_\mathrm{sf}$ and $l^T_\mathrm{sf}$ can be understood by noting that longitudinal conductivity is dominated by electrons with large group velocities, and spin-Hall conductivity is dominated by those with large anomalous velocities. The former electrons have a smaller spin-flip scattering cross-section on impurities compared to the latter, and hence the corresponding spin-diffusion length is longer. The results in Table~\ref{tab:coefs} also show that $l^T_\mathrm{sf}$ in all cases is approximately equal to or even less than the mean-free path. This behavior is consistent with the fact that the intrinsic spin-Hall conductivity tends to be dominated by Berry curvature ``hot spots'' in the Brillouin zone where spin-orbit coupling is comparable to the separation between the energy levels. It is natural to expect that scattering in and out of such hot spots has a high probability of spin flips \cite{fabian1998spin}, leading to $l^T_\mathrm{sf}\approx \lambda$. This feature, which may be generic for intrinsic spin-Hall transport, suggests that the widely used spin-diffusion model, which formally requires $l^T_\mathrm{sf}\gg \lambda$, may be inapplicable to this regime.

To gain insight in the role of interfacial scattering, we repeated the calculation at $E=E_F$ with $V_m=80$ mRy with spin-orbit coupling turned off in 15 ML of Pt next to each surface of the film. The resulting spin accumulation is shown in Fig.~\ref{fig:pt100sy}(a). Since spin-orbit coupling plays an essential role in both boundary condition terms in Eq.~(\ref{bc}), both $G_{sl}$ and $\sigma_\mathrm{SSH}$ become zero. The non-trivial boundary conditions are then located at layer 15 at the discontinuity in the electronic structure due to the turning off of the spin-orbit coupling. We discuss these boundary conditions at the end of Sec.~\ref{sec:current}. We note that when shifted by 15 ML toward the surface (red dashed line), the spin accumulation profile is quite close to its counterpart for the unmodified system. This similarity suggests that the spin accumulation profile is not particularly sensitive to the details of the electronic structure and disorder-induced scattering near the free surface.

\begin{table*}[hbt]
    \centering
    \begin{tabular}{c|c|c|S[table-format=2.1]|S[table-format=2.1]|S[table-format=1.2]|c|S[table-format=3.3]|S[table-format=3.2]|c|c|}
         \hline
         $E$ & $V_m$ (mRy) & $\rho$ (\unit{\micro\ohm\centi\meter}) & {$\lambda$ (nm)} & {$l^L_\mathrm{sf}$ (nm)} & {$l^T_\mathrm{sf}$ (nm)} & $\rho l^L_\mathrm{sf}$ (\SI{}{\femto\ohm\meter\squared}) & {$\theta^{(\mu)}_\mathrm{eff}$} & {$\sigma_\mathrm{SH}$} & $\theta_\mathrm{SH}$ & $\theta^{(j)}_\mathrm{eff}$\\
         \hline
         $E_F$ & 40  & 15.8  & 3.0 & 7.7 & 1.35 & 1.17  & 0.076     &  5.24 & 0.082   & 0.010 \\
         $E_F$ & 60  & 30.4  & 1.6 & 3.3 & 1.47 & 1.03  & 0.048     &  4.28 & 0.130   & 0.048 \\
         $E_F$ & 80  & 43.1  & 1.1 & 2.0 & 1.26 & 0.84  & 0.038     &  3.68 & 0.159   & 0.079 \\
         $E_F$ & 80 (S15) &  &     &     & 1.31 &       & 0.046     &  3.90 & 0.168   & 0.078 \\
         $E'$  & 40  & 4.50  & 10.2 & 32.2 & 7.0& 1.27  & -0.029    & -3.11 & $-0.014$ & $-0.028$  \\
         $E'$  & 60  & 10.3  & 4.4 & 14.1 & 3.3 & 1.47  & -0.033    & -2.32 & $-0.024$ & $-0.020$ \\
         $E'$  & 80  & 18.3  & 2.5 & 7.1 & 2.3  & 1.34  & -0.026    & -2.56 & $-0.047$ & $-0.026$\\
         $E'$  & 100 & 28.2  & 1.6 & 4.3 & 1.8  & 1.20  & -0.018    & -2.13 & $-0.060$ & $-0.031$ \\
         \hline
    \end{tabular}
    \caption{Parameters of the spin accumulation profile and the spin current in the 100-ML Pt film (1 ML = 0.196~nm). $\theta^{(\mu)}_\mathrm{eff}$ and $l^T_\mathrm{sf}$ are obtained by fitting $\mu_y(z)$ to Eq.~(\ref{muy2}). With that $l^T_\mathrm{sf}$ fixed, $\theta_\mathrm{SH}$ and $\theta^{(j)}_\mathrm{eff}$ are then obtained by fitting the spin current to Eq.~(\ref{jssol}). The row marked S15 corresponds to the fictitious system with spin-orbit coupling turned off in 15 ML near each surface. The units for $\sigma_\mathrm{SH}=\sigma\theta_\mathrm{SH}$ are $10^5$ \unit{\per\ohm\per\meter}. Details of the determination of the longitudinal spin-diffusion length are given in Appendix~\ref{sec:logitudinal} and details of the fitting procedure for other parameters are given in Appendix~\ref{sec:Fitting}. 
    }
    \label{tab:coefs}
\end{table*}

\subsection{Spin current}
\label{sec:current}

\begin{figure}[htb]
    \centering
    \includegraphics[width=0.9\columnwidth]{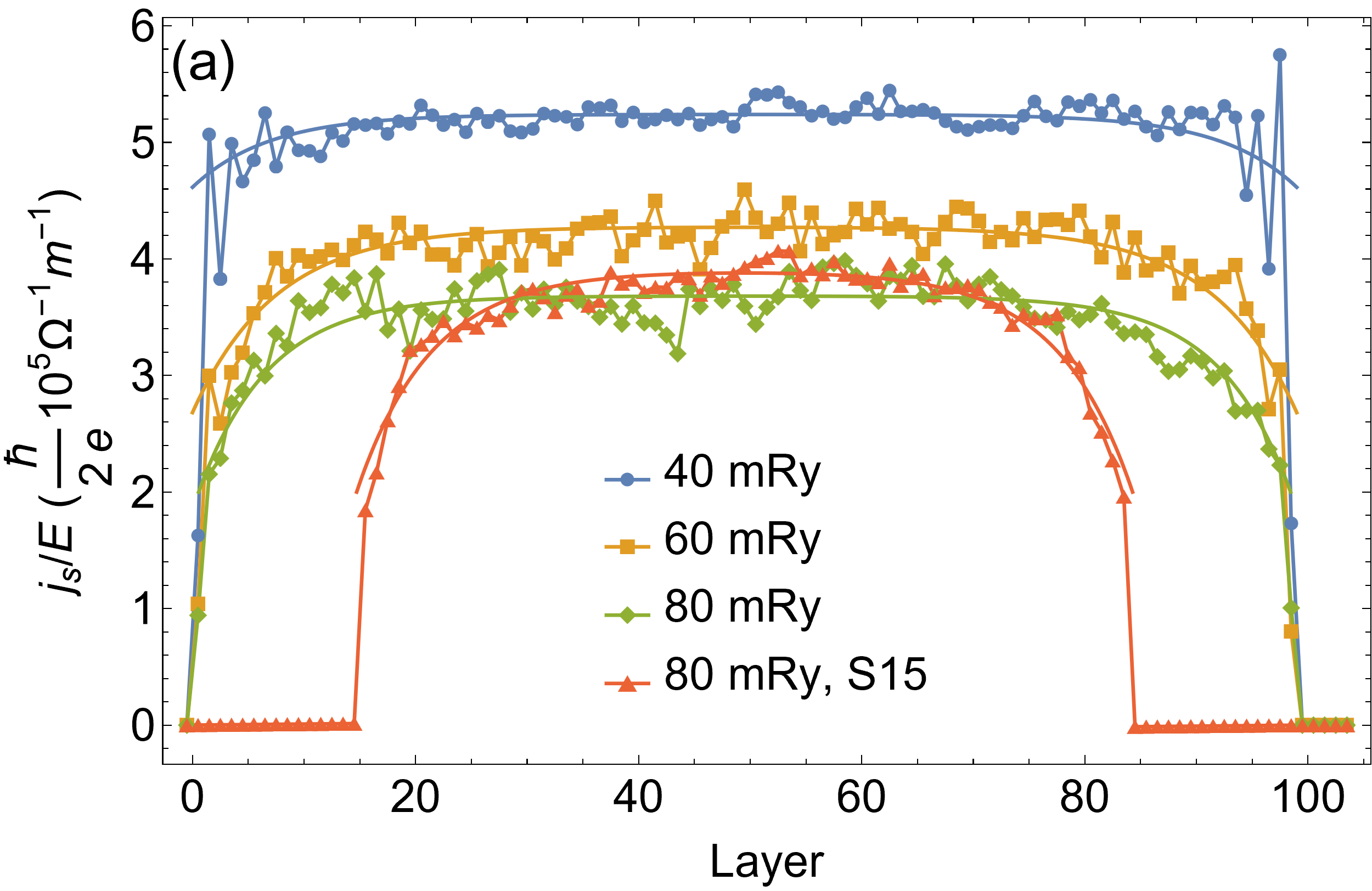}
    \includegraphics[width=0.9\columnwidth]{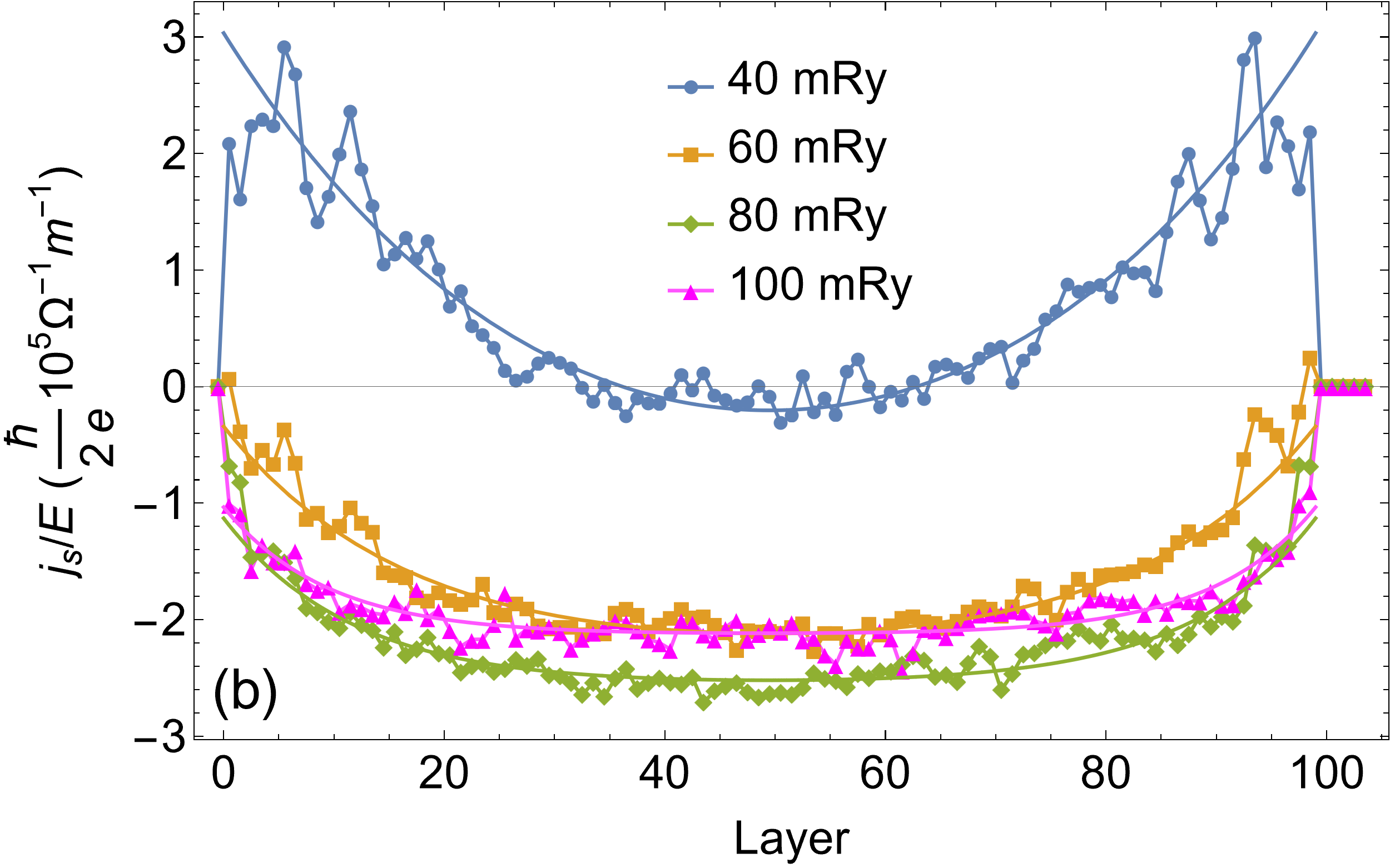}
    \caption{Spin current in the Pt film expressed as shown in Eq.~(12) in units of $10^5 \frac{\hbar}{2e}$ \unit{\per\ohm\per\meter}. The calculated intrinsic spin-Hall conductivity of Pt reported in the literature is close to $4\times10^5$ $\frac{\hbar}{2e}$
    \unit{\per\ohm\per\meter}. The smooth lines are fits to Eq.~(\ref{jssol}) where $l^T_\mathrm{sf}$ is taken from the spin accumulation fits.
    (a) At $E=E_F$. (b) At $E'=E_F-2.72$~eV.}
    \label{fig:jsypt100}
\end{figure}

Figure~\ref{fig:jsypt100}(a) shows the local conductivity $\sigma_{i,i+1}$ in the 100~ML Pt film at $E=E_F$ and Fig.~\ref{fig:jsypt100}(b) at $E=E_F-~2.72$~eV. We see that this quantity is approximately constant in the bulk at $E=E_F$, which is consistent with Fig.~\ref{fig:pt100sy}(a) showing that the thickness of the film is large compared with the transverse spin-diffusion length. On the other hand, finite curvature in the middle of Fig.~\ref{fig:jsypt100}(b), which is especially clear at $V_m=40$ mRy, is consistent with a longer transverse spin-diffusion length apparent in Fig.~\ref{fig:pt100sy}(b) at that energy.

\begin{figure}[htb]
    \centering
    \includegraphics[width=0.95\columnwidth]{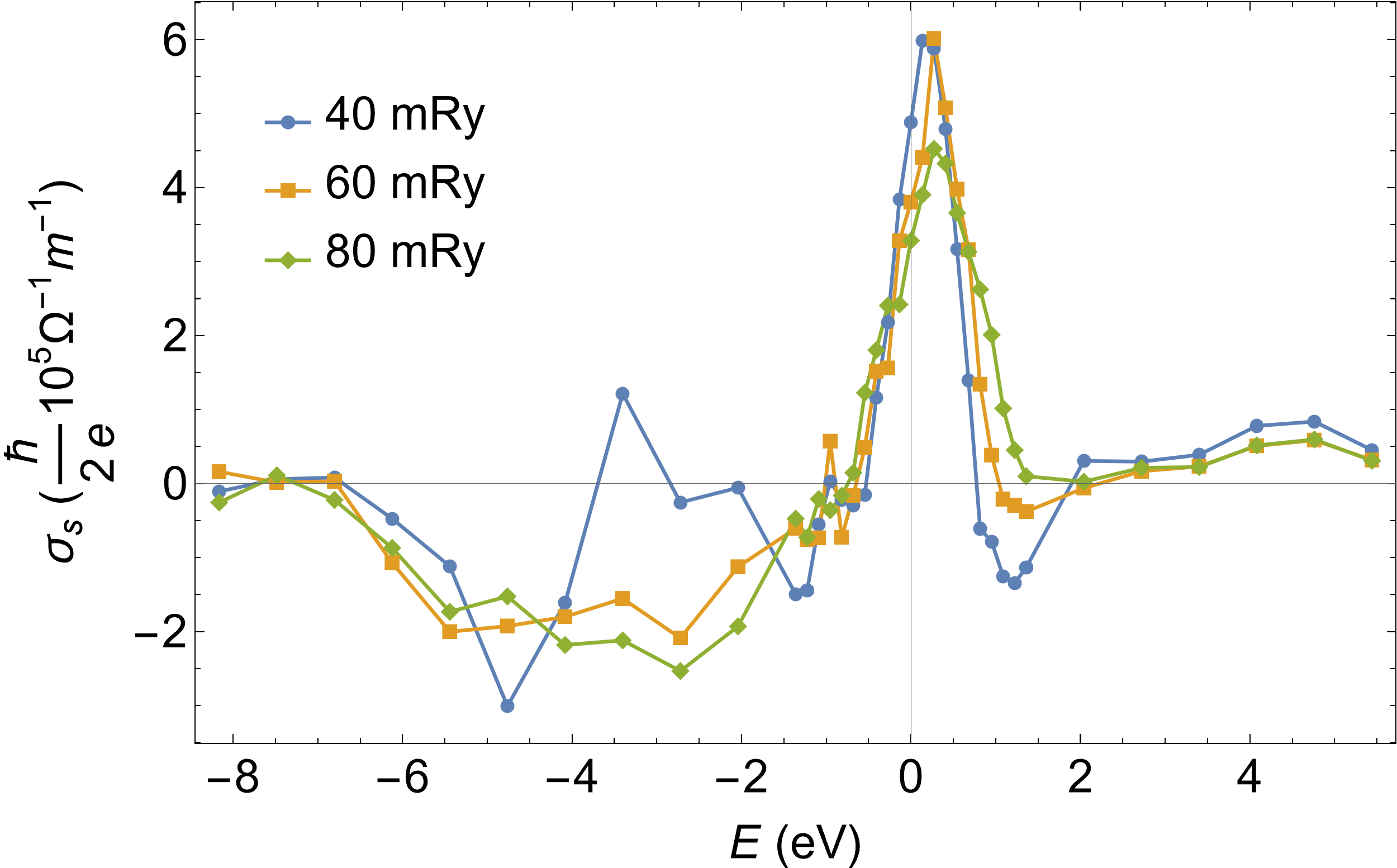}
    \caption{Spin conductivity $\sigma_c=(2e/\hbar)j_s(0)/E$ in the middle of the 100-ML Pt film as a function of the Fermi level for several disorder strengths $V_m$.}
    \label{fig:shcPt}
\end{figure}

We repeat the calculation of spin accumulation and spin currents in a wide range of energies, but with a smaller number of disorder samples. The results are too noisy to meaningfully extract a decay constant near the surface, but sufficient to compute the average of the local spin conductivity $\sigma_{i,i+1}$ over the central region ($41\leq i\leq 60$) of the film, shown in Fig.~\ref{fig:shcPt} for several values of disorder strength $V_m$. 
We denote this spin conductivity as $\sigma_c$. Within the spin-diffusion model, $\sigma_c=\sigma_\mathrm{SH}+\sigma_\mathrm{BF}(z)$ where $\sigma_\mathrm{BF}(z)$ is the backflow conductivity corresponding to the second term in (\ref{js}) and (\ref{jssol}). For energies below the physical Fermi energy, $\sigma_\mathrm{BF}(0)$ is only significant at energies between $-3.5$~eV and $-2$~eV with $V_m=40$ mRy; subtracting it from $\sigma_c$ reduces the ``bump'' at those energies but does not eliminate it.

At energies more than 1~eV above the physical Fermi energy, Eq.~(\ref{js}) formally gives a very large negative $\sigma_\mathrm{BF}(0)$ which is much larger than $\sigma_c$ shown in Fig.~\ref{fig:jsypt100}. However, we know from Fig.~\ref{fig:mfp} that the mean-free path exceeds the thickness of the film and even the length of the embedded supercell region in that energy range, and hence the results are not indicative of diffusive transport. Even if the length of the supercell were longer than the mean-free path, there is no basis for using the spin diffusion equation (\ref{js}) when the mean-free path exceeds the thickness of the film. Therefore, we assume that large $\sigma_\mathrm{BF}$ formally obtained at these energies are nonphysical. This is consistent with the physical expectation that the spin-Hall conductivity should rapidly decline as the Fermi level exits the $5d$ bands. Therefore, for simplicity we do not subtract $\sigma_\mathrm{BF}$ from $\sigma_c$ in Fig.~\ref{fig:shcPt} and treat $\sigma_c$ as a good representation of the spin-Hall conductivity, with the exception of $V_m=40$ mRy in the energy interval where that curve deviates from others for larger $V_m$.

As we see from Fig.~\ref{fig:shcPt}, the entire $\sigma_\mathrm{SH}(E)$ curve depends weakly on disorder strength in the chosen range of disorder strengths.
With the exception of the energies \qtyrange{2}{4}{eV} below the Fermi level at the low disorder level $V_m=40$ mRy, the magnitude and shape of the curve agree very well with the intrinsic contribution calculated using the Kubo formula \cite{Guo2008,salemi2022first}. The exception mentioned above is only partially accounted for by spin backflow, and there may be a positive extrinsic contribution at those energies at $V_m=40$ mRy. The sharp peak at the Fermi level is only slightly broadened and shifted by disorder while its height remains approximately constant. The small shift is likely due to our neglect of the Fermi level shift induced by disorder. This behavior indicates that the spin-Hall current in our diffusive system is dominated by the intrinsic contribution to the spin-Hall conductivity. 

For two energy points corresponding to $E_F$ and $E'$, the bulk and effective spin-Hall angles $\theta_\mathrm{SH}$ and $\theta^{(j)}_\mathrm{eff}$ are obtained by fitting the spin current to Eq.~(\ref{jssol}) with the $l^T_\mathrm{sf}$ parameter taken from the fit of the spin accumulation.
These parameters, along with the corresponding $\sigma_\mathrm{SH}$, are listed in Table~\ref{tab:coefs}.

At $E=E_F$, we see that the ``true'' spin-Hall angle $\theta_\mathrm{SH}$ and the effective spin-Hall angles obtained from the spin accumulation ($\theta^{(\mu)}_\mathrm{eff}$) and spin current ($\theta^{(j)}_\mathrm{eff}$) are all different. Moreover, while $\theta_\mathrm{SH}$ and $\theta^{(j)}_\mathrm{eff}$ increase with increasing disorder, $\theta^{(\mu)}_\mathrm{eff}$ decreases. These differences indicate that the spin-diffusion model is not valid in Pt at $E=E_F$. The situation is different at $E=E'$ where $\theta^{(\mu)}_\mathrm{eff}$ and $\theta^{(j)}_\mathrm{eff}$ are closer to each other but significantly different from $\theta_\mathrm{SH}$. The relationship, $\theta^{(j)}_\mathrm{eff}=\theta^{(\mu)}_\mathrm{eff}\neq\theta_\mathrm{SH}$, is captured by the spin-diffusion model, as seen from Eq.~(\ref{theff}) and discussed in Sec.~\ref{sec:SpinDiffusion}.

Since the data at $E=E'$ are not grossly inconsistent with the spin-diffusion model, we can ask if we can learn anything about the appropriate boundary conditions. The dependence of $\theta_\mathrm{eff}$ in Eq.~(\ref{theff}) on both $\theta_\mathrm{SSH}$ and $g_{sl}$ makes it impossible to determine their values separately. However, it is clear that the boundary condition from magnetoelectronic circuit theory \cite{brataas2006non}, i.e., the term with $G_{sl}$ in Eq.~(\ref{bc}), is not sufficient by itself without $\sigma_\mathrm{SSH}$. For example, at $V_m=40$ mRy we find $|\theta^{(j)}_\mathrm{eff}|\approx |\theta^{(\mu)}_\mathrm{eff}|>|\theta_\mathrm{SH}|$, which at $\sigma_\mathrm{SSH}=0$ would require a negative $g_{sl}\approx-0.5$. On the other hand, at $V_m=80$ mRy the relation is reversed, and we would have $g_{sl}\approx0.8$. Theoretically, $G_{sl}$ should be between zero and twice the Sharvin conductance and independent of the details of the bulk transport. 
These results suggest that the term with $\sigma_\mathrm{SSH}$ in the boundary condition is essential in this case.

The importance of $\sigma_\mathrm{SSH}$ is further borne out by the S15 results with the spin-orbit coupling turned off in the fifteen layers near the surfaces. In that calculation, there is no intrinsic spin Hall current in those fifteen layers because that spin current requires spin-orbit coupling. Further, there is no net diffusive spin current because the spin current must be zero at the outer interface and there is no spin relaxation without spin-orbit coupling. From the similarity in the full results and the results of the S15 calculation shifted by fifteen layers, it appears that the boundary conditions in Eq.~(\ref{bc}) depend more on the discontinuity in the spin-orbit coupling than on the existence of a surface \emph{per se}. In both cases, at the point where the intrinsic spin Hall conductivity changes from the bulk Pt value to zero, very similar spin accumulation profiles are created and diffuse similarly away. At both interfaces, the difference between the bulk spin Hall conductivity $\sigma_\mathrm{SH}$, which characterizes the flow of angular momentum toward the interface, and the value of the surface spin Hall conductivity $\sigma_\mathrm{SSH}$, which captures the flow of spin angular momentum into the lattice at the interface, gives the rate of spin accumulation. It appears that differences in these two values lead to extrinsic spin accumulation and diffusion whether the change in the electronic structure is substantial as at the surface or is relatively minor like where the spin-orbit coupling is turned off in the S15 calculation. One conclusion that can be drawn from the behavior of the spin accumulations and spin current in the S15 system is that the inverse spin-galvanic effect at the surface is not significant in this system.

\subsection{Discussion}
\label{sec:discuss}

Nair \emph{et al.}~\cite{nair2021spin} reported similar calculations for the spin current in a Pt film but came to different conclusions.
By taking into account the surface spin Hall effect and positing that the bulk spin Hall effect is reduced near the surface due to Fuchs-Sondheimer effects, they
find that the spin current, calculated with lattice disorder representing frozen phonons at room temperature, can be fit using the spin-diffusion model with a transverse spin-diffusion length equal to the longitudinal spin-diffusion length. We find that our results are not consistent with the spin-diffusion model if the spin accumulation is included in the analysis.

Although all of our results are consistent with an exponential decay away from the surface of both the spin accumulation and the deviations in the spin current from the value far from the surface, we find this decay constant is different than the longitudinal spin-diffusion length. Furthermore, the amplitudes we find for the spin accumulation and the excess spin current are not consistent with Eq.~(\ref{js}) and Eq.~(\ref{continuity}). For some disorder values, the spin current is larger than would be given by the diffusion of the spin density, and for other values it is smaller. These discrepancies may not hold for all cases, but they indicate violations of the standard semiclassical description of the spin-Hall effect in thin layers.

We note that Nair \emph{et al.}~\cite{nair2021spin} model thermal lattice disorder by displacing atoms from their equilibrium positions, which simulates phonon scattering characteristic of room temperature in single-crystal Pt. We use the Anderson disorder model and consider several stronger values of disorder leading to resistivities comparable to those in disordered thin Pt films typically used in spin-orbit torque measurements~\cite{manchon2019current}.

Our results have implications for the interpretation of experimental measurements in terms of other physical parameters.
We expect that measurements of spin-orbit torques as a function of the thickness of the heavy-metal layer will show an exponential dependence but caution against equating that decay constant with the usual spin-diffusion length. We note that the transverse spin-diffusion length $l^T_\mathrm{sf}$ extracted here from the spin accumulation profile is quite close to that found in the simulations of the thickness dependence of dampinglike torque in Co/Pt bilayers \cite{belashchenko2020interfacial} but significantly smaller than the usual longitudinal spin-diffusion length $l^L_\mathrm{sf}$.
It is also clear, from both our results and those of Nair \emph{et al.}~\cite{nair2021spin}, that the boundary conditions for intrinsic spin currents differ from those for diffusive spin currents as described by the magnetoelectronic circuit theory~\cite{brataas2006non}. However, our calculations cannot determine the correct boundary conditions, in part because of the violation of the semiclassical transport equations. In addition, the semiclassical separation of equations into bulklike regions connected by boundary conditions only makes sense if the spin-diffusion length is much larger than the microscopic scales, such as the Fermi wavelength and the thickness of the region where the electronic structure is modified by the presence of the interface. The discrepancies we find between the semiclassical description and our results indicate that such a separation is not valid.

Stamm \textit{et al}.~\cite{stamm2017magneto} have measured the transverse spin accumulation in Pt layers using the magneto-optical Kerr effect. While the interpretation of their measurement leads to spin Hall conductivities and spin Hall angles consistent with what we find, they find a transverse spin diffusion length that is much longer than we find and is consistent with the longitudinal spin diffusion length. The measured spin diffusion length is also a factor of two longer than that found by Nair \textit{et al}.~\cite{nair2021spin}. The measured longitudinal decay length is much longer than the decay length found in the Pt thickness dependence of measurements of the spin-orbit torque. We would expect that this decay length would be the same as the transverse spin diffusion length. Converting our surface spin accumulation to a spin density gives a result that is roughly a factor of six smaller than that extracted from the measurements of Stamm \textit{et al}.~\cite{stamm2017magneto}. We have no explanation for these discrepancies.

Most numerical calculations of the spin-Hall effect~\cite{Guo2008,tanaka2008intrinsic,salemi2022first} are based on the Kubo formalism with a constant broadening approximation. It is useful to compare and contrast that approach with the present one.  Within the Kubo formalism, there is a clear distinction between interband and intraband contributions to the response for typically used values of the broadening parameter. Intraband contributions are derived from the response of individual eigenstates at the Fermi energy, whereas interband contributions involve matrix elements between occupied and unoccupied states. For experimentally relevant electron lifetimes in the Kubo formalism, the spin Hall current is dominated by the intrinsic contribution.
Our NEGF approach with explicit disorder averaging includes intraband and interband contributions on the same footing, but the results appear to be dominated by non-equilibrium changes in occupation of disordered states at the Fermi energy. We note that a Kubo approach which includes vertex corrections also lacks a clear separation into interband and intraband contributions, as the vertex corrections contain interband contributions~\cite{PhysRevLett.105.036601}.

It is intriguing that within the statistical uncertainty of our calculations, the calculated spin current in the middle of the slab agrees with the value found within the Kubo formalism for the intrinsic contribution.
A possible interpretation of the connection between the two calculations is that the matrix elements in the intrinsic calculations are like susceptibilities, which capture the extent to which the wave function will change in the presence of disorder that mixes bands and locally breaks inversion symmetry.

Within the Kubo formalism, Ref.~\cite{go2020theory} derives a spin continuity equation, which relates the interband spin current divergence and spin-orbital torque to the time derivative of the intraband spin accumulation.
This analysis shows that the intrinsic quantities do not satisfy a steady-state spin continuity equation on their own, and there must be a conversion between intrinsic and extrinsic ``channels'' in the vicinity of interfaces and other inhomogeneities. In fact, the intrinsic spin accumulation vanishes in a non-magnetic system due to time-reversal symmetry, while mechanisms for generation of extrinsic spin currents in the vicinity of an interface have been discussed in the literature~\cite{PhysRevB.76.085319}.
Without including vertex corrections in lieu of an explicit treatment of disorder, the Kubo formalism does not properly describe the diffusion of intraband spin accumulation \cite{rammer1986quantum}, so that it does not quantitatively capture the behavior found in our calculations.
On the other hand, in calculations with explicit disorder averaging the spin current divergence is equal and opposite to the spin-orbital torque as indicated by Eq.~(\ref{eq:dsdt}), and we cannot distinguish between intrinsic spin currents due to band structure effects and extrinsic spin currents due to the diffusion of accumulated spins. 

Interconversion between intrinsic and extrinsic components near the interface is likely to result in an additional length scale associated with intrinsic effects, which is not included in Eqs.~\eqref{js} and \eqref{continuity}. This feature may present one of the mechanisms through which the conventional spin diffusion model is violated.

\section{Orbital accumulation}
\label{sec:orbital}

Recent work~\cite{tanaka2008intrinsic,go2020orbital} suggests that currents of orbital moments play an important role in spin-Hall effects and angular momentum transfer from heavy metals to the magnetizations of ferromagnets. Motivated by this, we next consider orbital accumulation present in this system.

To compare current-induced orbital and spin densities in the same units, we define orbital accumulation as $\boldsymbol{\mu}_\mathrm{orb}=\mathop\mathrm{Tr} \hat{\mathbf{l}} \hat\rho/{eN_s}$, where $\hat\rho$ is the on-site density matrix and $\mathbf{l}$ the angular momentum operator divided by $\hbar$ \footnote{It is common to refer to spin accumulation as a synonym to the spin potential. We emphasize that there is no clear justification to introduce an ``orbital potential'' by analogy with the spin potential, which owes its existence to the weak coupling between two spin channels.}. Only the $y$ component is allowed by symmetry in the case at hand. 

\begin{figure}
    \centering
    \includegraphics[width=0.9\columnwidth]{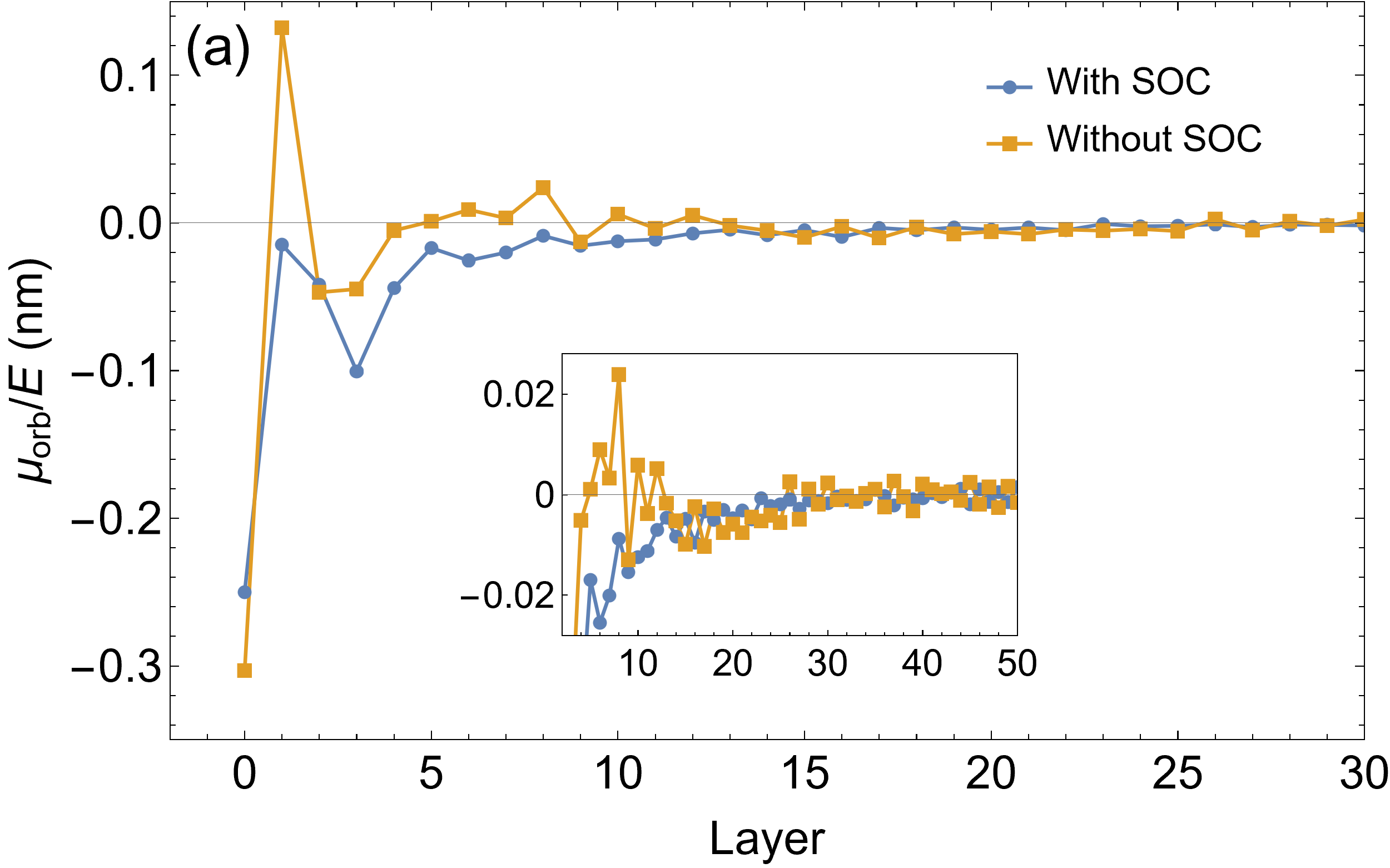}
    \includegraphics[width=0.9\columnwidth]{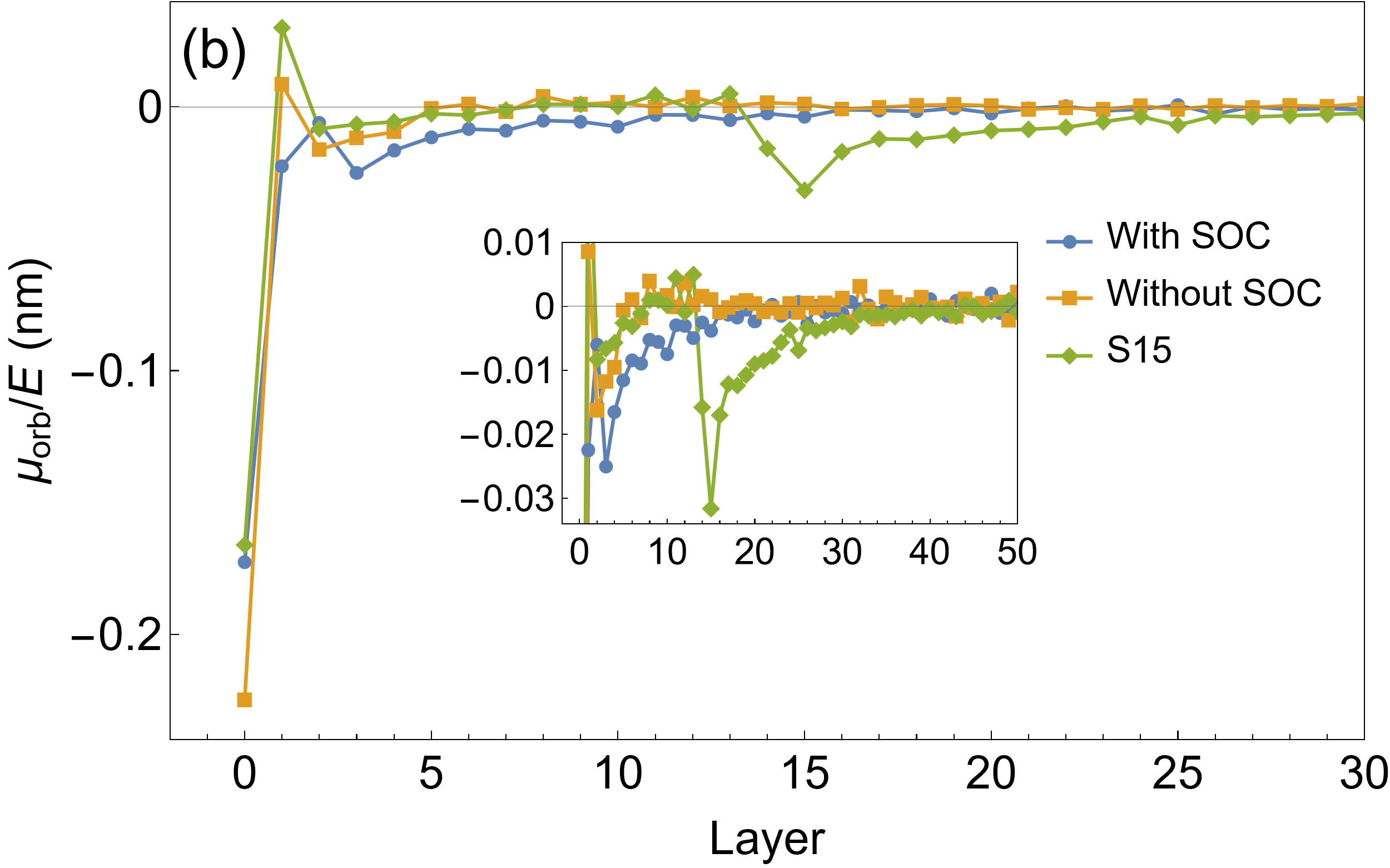}
    \includegraphics[width=0.85\columnwidth]{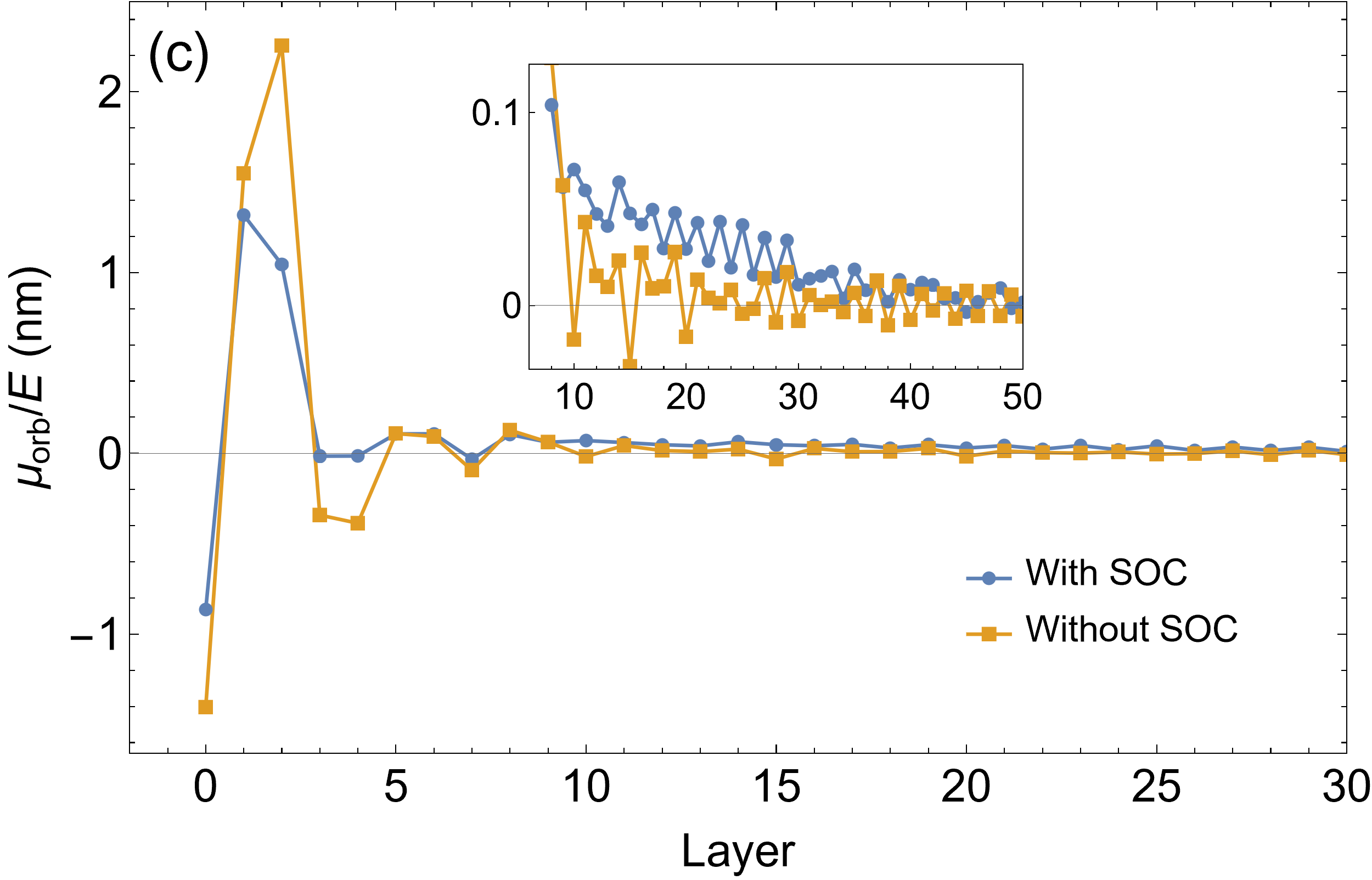}
    \caption{Orbital accumulation profiles $\mu_\mathrm{orb}(z)/E$ in a 100-ML Pt film with (a) $V_m=40$ mRy at $E=E_F$, (b) $V_m=80$ mRy at $E=E_F$, and (c) $V_m=40$ mRy at $E=E'$ calculated with or without spin-orbit coupling (SOC). Panel (b) includes the data for the S15 system which has spin-orbit coupling turned off in 15 monolayers of Pt near each surface. Insets show the same data on smaller scales.}
    \label{fig:pt100oy}
\end{figure}

Figure~\ref{fig:pt100oy} shows the orbital accumulation profiles in a 100-ML Pt film calculated with or without spin-orbit coupling. Panels (a) and (b) are for transport at $E=E_F$ with $V_m=40$~mRy or 80~mRy, respectively, and panel (c) is for $E=E'$ with $V_m=40$ mRy. Panel~(b) also includes the orbital accumulation for the system (S15) with spin-orbit coupling turned off in the 15 layers closest to the surface.
In contrast with the spin accumulation profiles shown in Fig.~\ref{fig:pt100sy}, the orbital accumulations are large only in (at most) a few surface layers. Just a few monolayers into the bulk of the film, the orbital accumulation becomes more than an order of magnitude smaller. Thus, orbital accumulation is largely a surface effect.

First, let us focus on the behavior of the relatively small orbital accumulation a few monolayers away from the surface. If spin-orbit coupling is included, we observe a decaying smooth orbital accumulation profile. However, this slowly decaying orbital accumulation is either absent or strongly suppressed when spin-orbit coupling is turned off.
This suggests that in our system the phenomenology of spin diffusion has no analog in the orbital sector due to strong nonconservation of orbital momentum associated with interatomic hopping and the presence of short-range disorder. Indeed, because a scattering event resulting in large momentum transfer is also expected to scramble the orbital momentum, the characteristic decay length for orbital accumulation in a system with localized disorder should be similar to the mean-free path.
On the other hand, the small orbital accumulation observed only in the presence of spin-orbit coupling arises as a second-order effect induced by the spin accumulation. This picture provides a plausible interpretation for the unexpectedly long length scales observed in recent experiments on sputtered Ni$_{81}$Fe$_{19}$/W(Ti) and Ni/W(Ti) films~\cite{Hayashi2023}. Our results also suggest that consideration of disorder beyond the broadening approximation may be important for analyzing orbital transport effects~\cite{tanaka2008intrinsic,go2020orbital,Lee2021,Hayashi2023}.

The coupling between orbital moments and lattice does not allow us to obtain the orbital Hall current in the same way we compute the spin Hall current. In general, spins are coupled to orbital moments and any magnetization, which is zero in the non-magnetic system we consider here. Thus, the spin current can be calculated from the coupling to the orbital moments as in Eq.~(\ref{eq:spincurrentdifference}). On the other hand, a similar procedure does not work for orbital currents because they not only couple to the spins, but also couple strongly to the lattice through the crystal field potential when they hop between sites. Computing the orbital current then requires computing the coupling to the lattice as well as the coupling to the spins. 

Strong coupling of orbital moments to the lattice explains why we can deduce that orbital Hall currents do not play an important role in the boundary conditions for the spin accumulation and spin current. This deduction is based on turning spin-orbit coupling off in the layers near the surface, see the S15 curves in Fig.~\ref{fig:pt100sy}, Fig.~\ref{fig:jsypt100} and Fig.~\ref{fig:pt100oy}. Disregarding this S15 calculation, one possible interpretation of our results could be the following. An intrinsic orbital Hall current leads to a large orbital Hall accumulation near the surface. Although this accumulation does not diffuse significantly away from the surface, it could, through spin-orbit coupling, create the spin accumulation and diffusive spin current we observe. This conjecture requires a spatial overlap between the orbital accumulation and spin-orbit coupling. In the S15 calculation, the spin-orbit coupling is turned off at the surface where there is substantial orbital accumulation, so that orbital accumulation cannot be playing a significant role in the boundary conditions between the spin current and orbital accumulation.

There is an increase in orbital accumulation near layer 15 in the S15 calculation. We attribute this peak to the discontinuity in the electronic structure due to abruptly turning off the spin-orbit coupling as discussed in Sec.~\ref{sec:discuss}. However, this peak is smaller than the peak at the surface, which is substantially the same as it is in the presence of spin-orbit coupling, suggesting that the surface orbital accumulation is  associated primarily with the effect of the surface-modified  potential rather than with spin-orbit coupling.

\section{Conclusions}
\label{sec:conclusions}

Our analysis of the spin accumulation and spin current in a disordered Pt film suggests the following conclusions. First, the results reveal significant deviations from the predictions of the spin diffusion theory, even with the modified boundary condition appropriate for intrinsic spin current generation. The key indicator of these deviations is the difference in the effective spin-Hall angles $\theta^{(\mu)}_\mathrm{eff}$ and $\theta^{(j)}_\mathrm{eff}$ extracted, respectively, from the spin accumulation and spin current profiles, and the different trends exhibited by these quantities as a function of disorder strength (see Table~\ref{tab:coefs}). We suggest the failure of the spin-diffusion model in Pt is due to the effective spin-diffusion length being too short and comparable to the mean-free path.

Notably, the results for a hypothetical system obtained by moving the Fermi level in Pt down by 2.72~eV are consistent with the spin-diffusion model with a modified boundary condition (\ref{bc}) reflecting the reabsorption of the spin current into the lattice near the surface. Even here, however, the spin-diffusion length for transverse transport is a few times shorter than the conventional one for longitudinal transport. This difference may be understood by noting that the spin-Hall current is likely dominated by low-velocity states near spin-orbit ``hot spots,'' while longitudinal transport is dominated by high-velocity states. 

Orbital accumulation in the Pt film is strongly localized near the surfaces. Only in the presence of spin-orbit coupling does it acquire a small component penetrating deeper into the bulk, which may be interpreted as a second-order effect induced by the spin accumulation. This result is consistent with the effective relaxation length for orbital accumulation being on the order of the mean-free path in systems with localized disorder.

Although the present paper is concerned with a nonmagnetic Pt film, our results have implications for the description of spin-orbit torques in bilayers. Violation of the spin-diffusion equations in Pt should carry over to any heterostructure that contains a Pt layer. If the spin-diffusion equations are satisfied, a distinction should generally be made between the transverse and longitudinal spin-diffusion lengths $l^T_\mathrm{sf}$ and $l^L_\mathrm{sf}$. The characteristic length scale on which the spin-orbit torque reaches its asymptotic limit as a function of the heavy-metal thickness is expected to be associated with $l^T_\mathrm{sf}$, which is approximately confirmed by the comparison of the spin-diffusion lengths listed in Table~\ref{tab:coefs} with those obtained from the torque calculations \cite{belashchenko2020interfacial}. Finally, the term with $\sigma_\mathrm{SSH}$ in the boundary condition (\ref{bc}) is likely generic for intrinsic spin current and also important at the interface with a ferromagnet, in addition to the term with $G_{sl}$ which turns into a (complex) spin-mixing conductance there~\cite{Baez2020}. Therefore, fitting of the experimental data for the spin-orbit torque to, say, Eq.~(\ref{fig:Thetaeff}) can give incorrect, and possibly unphysical, results for the spin-mixing conductance if $\sigma_\mathrm{SSH}$ is neglected.

\acknowledgements

We are grateful to Xin Fan for useful discussions and to Paul Kelly for his critical remarks about the manuscript.
KDB, GGBF, and WF were supported by the National Science Foundation through Grant No. DMR-1916275. AAK was supported by the U.S. Department of Energy, Office of Science, Basic Energy Sciences, under Award No. DE-SC0021019. M. van Schilfgaarde was supported by National Renewable Energy Laboratory, operated by Alliance for Sustainable Energy, LLC, for the U.S. Department of Energy, Office of Science, Basic Energy Sciences, Division of Materials, under Contract No. DE-AC36-08GO28308. Calculations were performed utilizing the Holland Computing Center of the University of Nebraska, which receives support from the Nebraska Research Initiative.

\appendix

\section{Fermi sea contribution to the spin current}
\label{Fermi-sea}

Here we estimate the Fermi-sea contribution to the transverse spin current in a disordered Pt film. We consider a 50-ML-thick film and use a $100\times 1$ supercell, averaging over 25 disorder configurations with $V_m=40$~mRy or 80~mRy. A bias of 0.136~mV is applied symmetrically as a linear potential across the device. For each disorder configuration, the spin-orbital torques at a positive and negative bias are subtracted to obtain the bias derivative. This procedure removes disorder-dependent equilibrium torque and improves the convergence of the disorder averages. The Fermi-sea term involves integration over the filled states,
which is usually performed using a semi-elliptical contour for equilibrium quantities \cite{pashov2020questaal}. Here we shift the semi-ellipse by 1~mRy along the imaginary axis and add vertical segments connecting its ends to the real axis. Legendre quadrature with 101 points is used on the semi-ellipse, and a uniform 10-point mesh on each vertical segment.

\begin{figure}
    \centering
    \includegraphics[width=0.9\columnwidth]{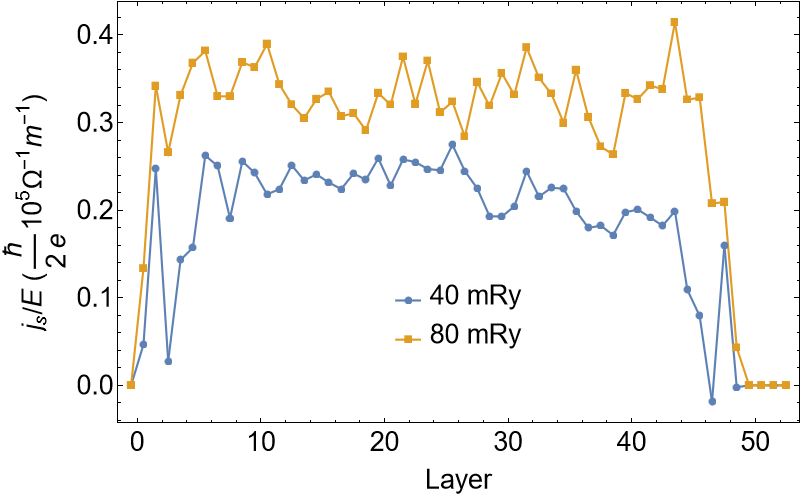}
    \caption{Fermi-sea contribution to spin current in a 50-ML Pt film at $V_m=40$ and 80~mRy at $E=E_F$.}
    \label{fig:Fermisea}
\end{figure}

Figure~\ref{fig:Fermisea} shows the Fermi-sea contribution to spin current in the 50-ML Pt film at $V_m=40$~mRy and 80~mRy. We see that this contribution is only about $2\times10^4 (\hbar/2e)$ $\unit{\per\ohm\per\meter}$ in the middle of the film, which is less than 10~\% of the Fermi-surface contribution shown in Fig.~\ref{fig:jsypt100}.

\section{Longitudinal spin-diffusion length}
\label{sec:logitudinal}

In order to determine the spin-diffusion lengths for longitudinal transport, we calculate the spin-conserving $G_{\uparrow\uparrow}$ and spin-flip $G_{\uparrow\downarrow}$ conductances for a disordered Pt supercell embedded between semi-infinite ballistic Pt leads and fit the results to the predictions of the magnetoelectronic circuit theory \cite{SML2016,Baez2020}. This fit is performed as follows. Within the circuit theory, the properties of a nonmagnetic layer are described by the conductances $G_{cc}$, $\mathcal{G}_{ss}$, and $\mathcal{G}_m$ defined as \cite{Baez2020}
$G_{cc}=2(G_{\uparrow\uparrow}+G_{\uparrow\downarrow})$, $\mathcal{G}_{ss}=2(G_{\uparrow\uparrow}-G_{\uparrow\downarrow})$ and $\mathcal{G}_{m}=4(G_{\uparrow\downarrow}+G^R_{\uparrow\downarrow})$, where $G^R_{\uparrow\downarrow}$ is the conductance corresponding to spin-flip reflection from the given layer. These conductances are defined with respect to voltages applied between fictitious nodes and are not directly comparable with the Landauer-B\"uttiker (LB) calculations of the disordered layer embedded between ballistic leads.

First, because we are dealing with relatively thin layers whose conductance is not very small compared to the Sharvin conductance $G_\mathrm{SH}$, the conductances need to be renormalized \cite{renormalization} as follows \cite{SML2016}: $\tilde G_{a}^{-1}=G_{a}^{-1}-G_\mathrm{SH}^{-1}$, where $G_{a}$ stands for $G_{cc}$, $\mathcal{G}_{ss}+\mathcal{G}_m$, or $\mathcal{G}_m/2$. Note that $\mathcal{G}_{ss}$ and $\mathcal{G}_m$ are renormalized as a group but independent of $G_{cc}$.
The thickness dependence of the renormalized conductances in the diffusive regime is known \cite{Baez2020}:
\begin{align}
    \tilde{G}_{cc}&=\frac{\sigma A}{d} ,\label{Gcc}\\
    \tilde{\mathcal{G}}_{ss}&=\tilde{G}_{cc}\frac{\delta_N}{\sinh \delta_N} ,\label{Gss}\\
    \tilde{\mathcal{G}}_{m}&=\tilde{G}_{cc}\delta_N\tanh \frac{\delta_N}{2}\label{Gm},
\end{align}
where $\delta_N=d/l^L_\mathrm{sf}$. Because the relevant range of thicknesses is up to a few spin-diffusion lengths, renormalization is important if the parameter $\kappa_N=\sigma A/(G_\mathrm{SH}l^L_\mathrm{sf})\sim \lambda/l^L_\mathrm{sf}$ is not small, where $\lambda$ is the mean-free path.

Second, because there can be substantial scattering at the embedding planes where the disordered Pt supercell is attached to ballistic leads, we introduce additional layers on both sides of the active region to capture this scattering, which is described by unknown renormalized conductances. As an approximation, they are described by the same equations (\ref{Gcc})--(\ref{Gm}) parameterized by $\delta_\mathrm{int}$ and $\kappa_\mathrm{int}$. The conductances of the diffusive region and of the two interfacial layers on each side are concatenated as prescribed by the circuit theory. ``Undoing'' the renormalization gives the conductances $G^\mathrm{LB}_{cc}$ and $\mathcal{G}^\mathrm{LB}_{ss}$ for the whole system which can be directly compared with the results of first-principles LB calculations.

The LB resistance $R_\mathrm{LB}=1/G^\mathrm{LB}_{cc}$ of the whole system is $R_\mathrm{LB}G_\mathrm{SH}=1+\delta_N/\kappa_N+2\delta_\mathrm{int}/\kappa_\mathrm{int}$. Fitting the calculated $R_\mathrm{LB}(d)$ to a straight line fixes $\delta_\mathrm{int}/\kappa_\mathrm{int}$ and $l^L_\mathrm{sf}\kappa_N$. Then we fit $\log \mathcal{G}^\mathrm{LB}_{ss}$ to the prediction of the circuit theory, fixing the remaining parameters $l^L_\mathrm{sf}$ and $\kappa_\mathrm{int}$.

\begin{figure*}
    \centering
    \includegraphics[width=0.85\columnwidth]{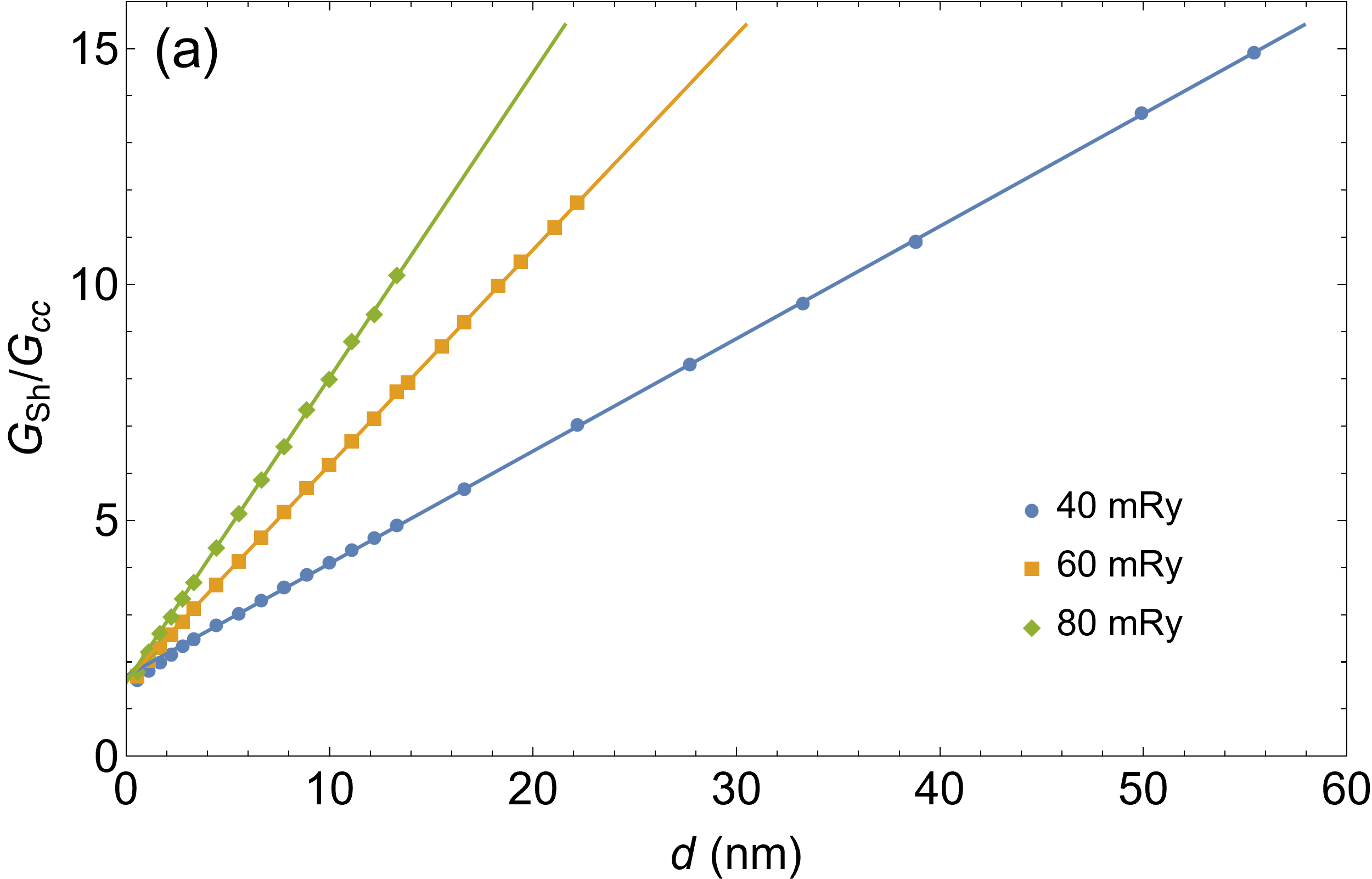}\hskip2em
    \includegraphics[width=0.85\columnwidth]{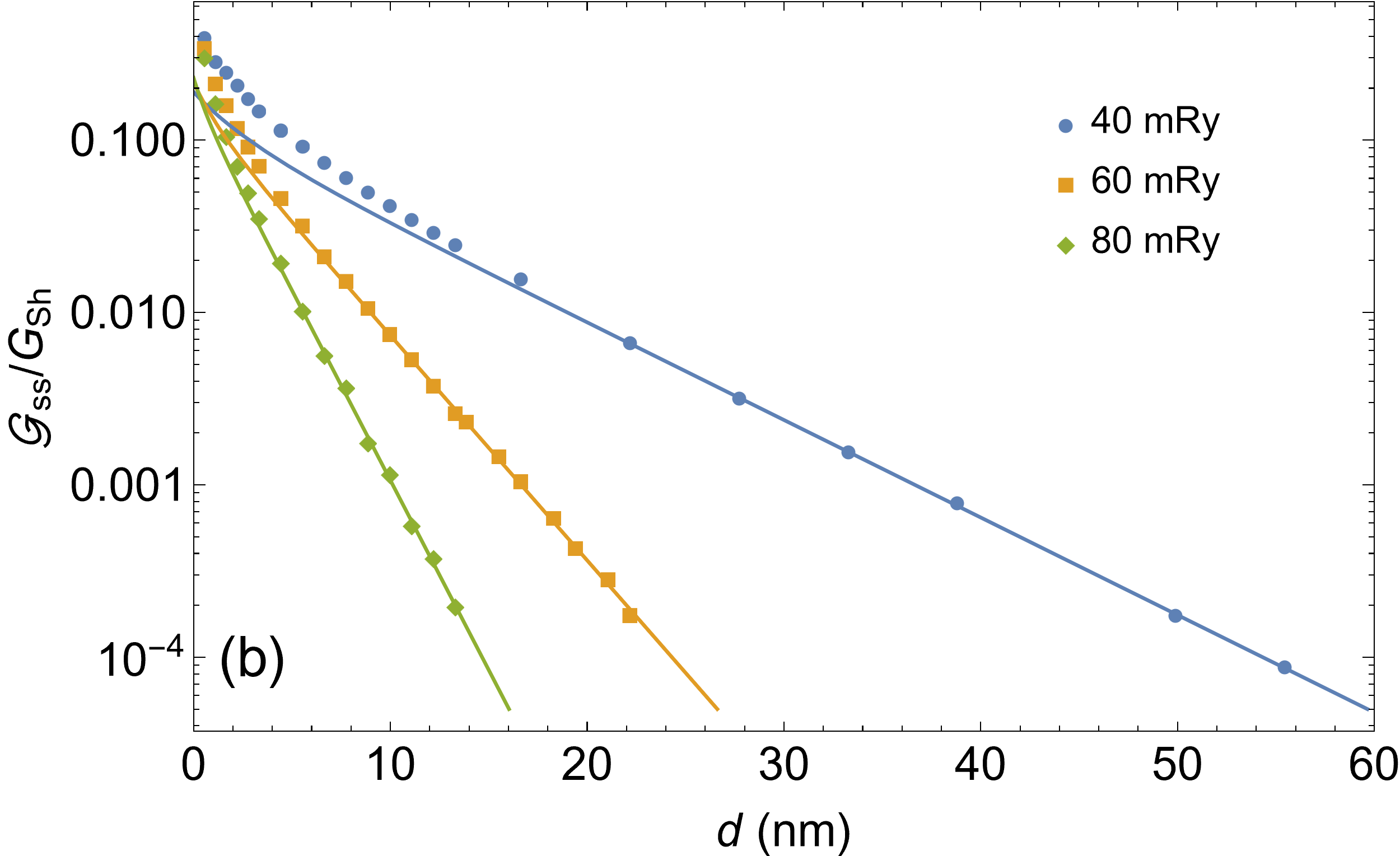}
    \caption{Fits of the calculated (a) reduced resistance $G_\mathrm{SH}/G_{cc}$ and (b)the reduced spin conductance $\mathcal{G}_{ss}/G_\mathrm{SH}$ as a function of the layer thickness $d$ in bulk Pt calculated at $E_F$ using the $16\times1$ lateral supercell.}
    \label{fig:lsflfits}
\end{figure*}

Unfortunately, the inclusion of renormalization and fictitious interface regions does not significantly improve the fits to the data at small thicknesses, which is due to the violation of the assumptions of the circuit theory as $d$ becomes comparable to $\lambda$.  Therefore, we only include points with $d>6\lambda$ in the fitted datasets, making sure that the values of $\mathcal{G}^\mathrm{LB}_{ss}$ are reliably averaged over disorder. The resulting fits are shown in Fig.~\ref{fig:lsflfits}.

The fitted values of $l^L_\mathrm{sf}$ are listed in Table~\ref{tab:coefs}. We see that the disorder dependence of $l^L_\mathrm{sf}$ approximately follows the expected Elliott-Yafet behavior that $\rho l^L_\mathrm{sf}$ is approximately constant at both $E_F$ and $E'$, although at $E_F$ the decline of 
$l^L_\mathrm{sf}$ with increasing disorder is somewhat faster than $\rho^{-1}$. The $\rho l^L_\mathrm{sf}$ product at $E_F$ of about \SI{1}{\femto\ohm\meter\squared} is notably larger than that reported by Wesselink \emph{et al.} \cite{Wesselink2019} for a more realistic frozen-phonon disorder model. The difference is not unexpected because, in the Elliott-Yafet mechanism, the spin-flip to spin-conserving scattering ratio depends on the nature of the impurity.

\section{Fitting to Semiclassical Models}
\label{sec:Fitting}

In this appendix we show the fits used to determine the values given in Table~\ref{tab:coefs} and used to draw our quantitative conclusions. We use Eq.~(\ref{muy2}) to compare the simulated spin density with the spin-diffusion model
\begin{align}
    [\Delta^{(\mu)}]^2=& \frac{1}{d-2\epsilon} 
    \int\limits_{-d/2+\epsilon}^{d/2-\epsilon} dz \nonumber\\&
    \left( \frac{\mu_y(z)}{2El_\mathrm{sf}}-\theta_\mathrm{eff}^{(\mu)} \frac{\sinh{z/l_\mathrm{sf}}}{\cosh{d/l_\mathrm{sf}}}\right)^2  ,
    \label{eq:fitmu}
\end{align}
and Eq.~(\ref{jssol}) to compare the simulated spin current with the spin-diffusion model
\begin{align}
    [\Delta^{(j)}]^2=& \frac{1}{d-2\epsilon} \int\limits_{-d/2+\epsilon}^{d/2-\epsilon} dz \nonumber\\&
    \left( \frac{\bar{\jmath}_s(z)}{j} - \theta_\mathrm{SH}+\theta_\mathrm{eff}^{(j)}\frac{\cosh{z/l_\mathrm{sf}}}{\cosh{d/l_\mathrm{sf}}} \right)^2  ,
    \label{eq:fitj}
\end{align}
where $j=\sigma E= E/\rho$ is the longitudinal current and $\epsilon$ represents the points near the surface that are omitted from the fits. Generally, $l_\mathrm{sf}$, $\theta_\mathrm{SH}$, $\theta_\mathrm{eff}^{(\mu)}$, and $\theta_\mathrm{eff}^{(j)}$ are treated as fitting parameters, but in some cases, they are frozen at values fixed by other fits. We find that as long as the two layers closest to the surface are omitted from the fit, the fits are insensitive to exactly how many are omitted.

\begin{figure*}
    \centering
    \includegraphics[width=1.7\columnwidth]{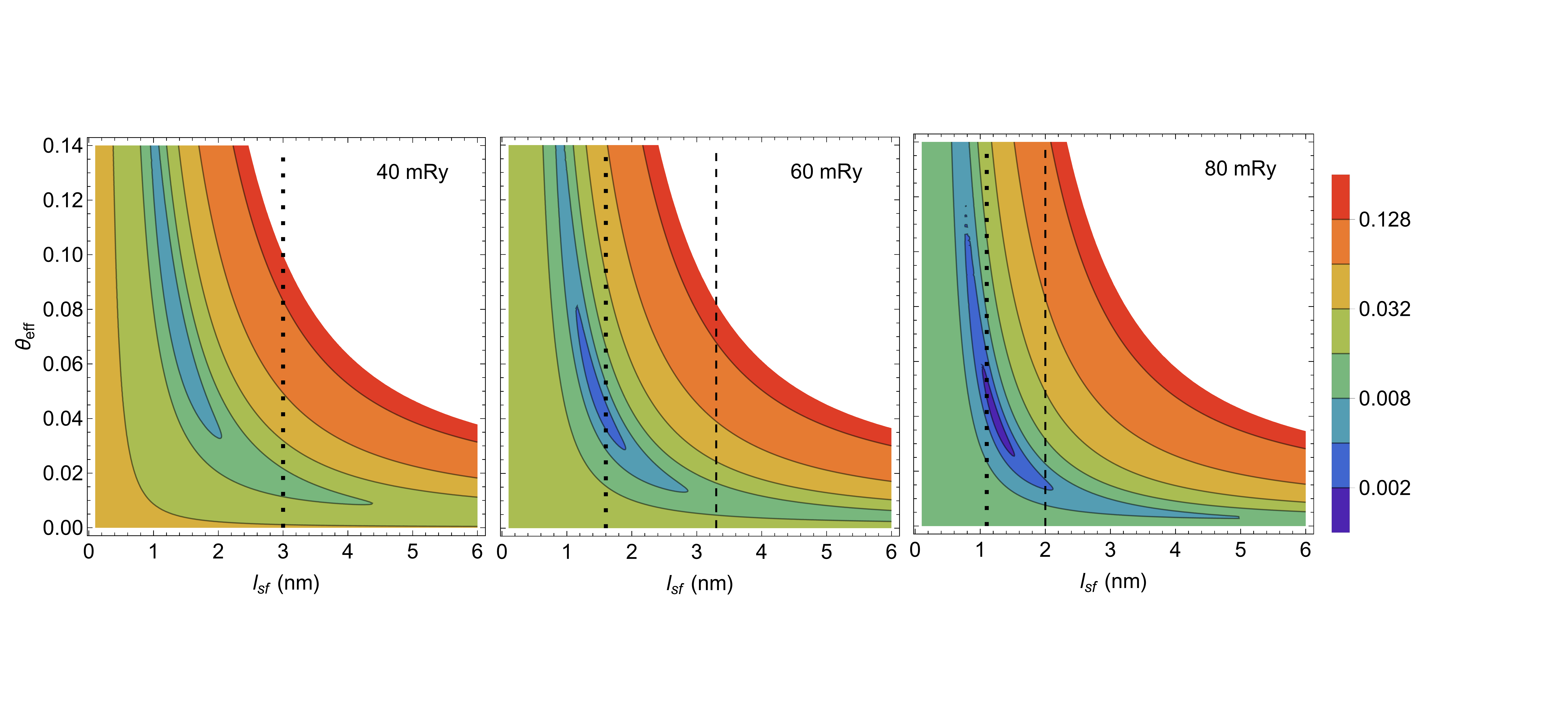}
    \caption{Fitting error $\Delta^{(\mu)}$ for the spin accumulation, as a function of $l_\mathrm{sf}$ and $\theta_\mathrm{eff}$ at $E_{F}$ for three different values of disorder. The dotted line indicate the mean free path for each value of disorder and the dashed lines indicate the longitudinal spin-diffusion length as discussed in Appendix~\ref{sec:logitudinal}. For 40~mRy disorder, the longitudinal spin-diffusion length is off the scale.}
    \label{fig:ContourPlots}
\end{figure*}

Fig.~\ref{fig:ContourPlots} shows the fits to the spin accumulation data shown in Fig.~\ref{fig:pt100sy}(a) fit using Eq.~(\ref{eq:fitmu}). The two fitting parameters are the transverse spin-diffusion length $l_\mathrm{eff}^{T}$ and the amplitude of the exponential spin accumulation near the interfaces, $\theta_\mathrm{eff}^{(\mu)}$. The fits tightly constrain the spin-diffusion length and more weakly constrain the amplitude. Also indicated on the plot are the longitudinal spin-diffusion lengths discussed in Appendix~\ref{sec:logitudinal}. These values are clearly inconsistent with the fits to the interfacial spin accumulation. 

Also indicated on the panels are the mean free paths as found from Fig.~\ref{fig:mfp}. The semiclassical model for spin diffusion is based on the approximation that the spin-diffusion length is long compared to the mean free path. The closeness of the transverse spin-diffusion lengths and the mean free paths and the differences between the longitudinal and transverse spin-diffusion length indicate that the spin-diffusion model might not be appropriate for the system. We give further evidence below.

\begin{figure}
    \centering
    \includegraphics[width=0.85\columnwidth]{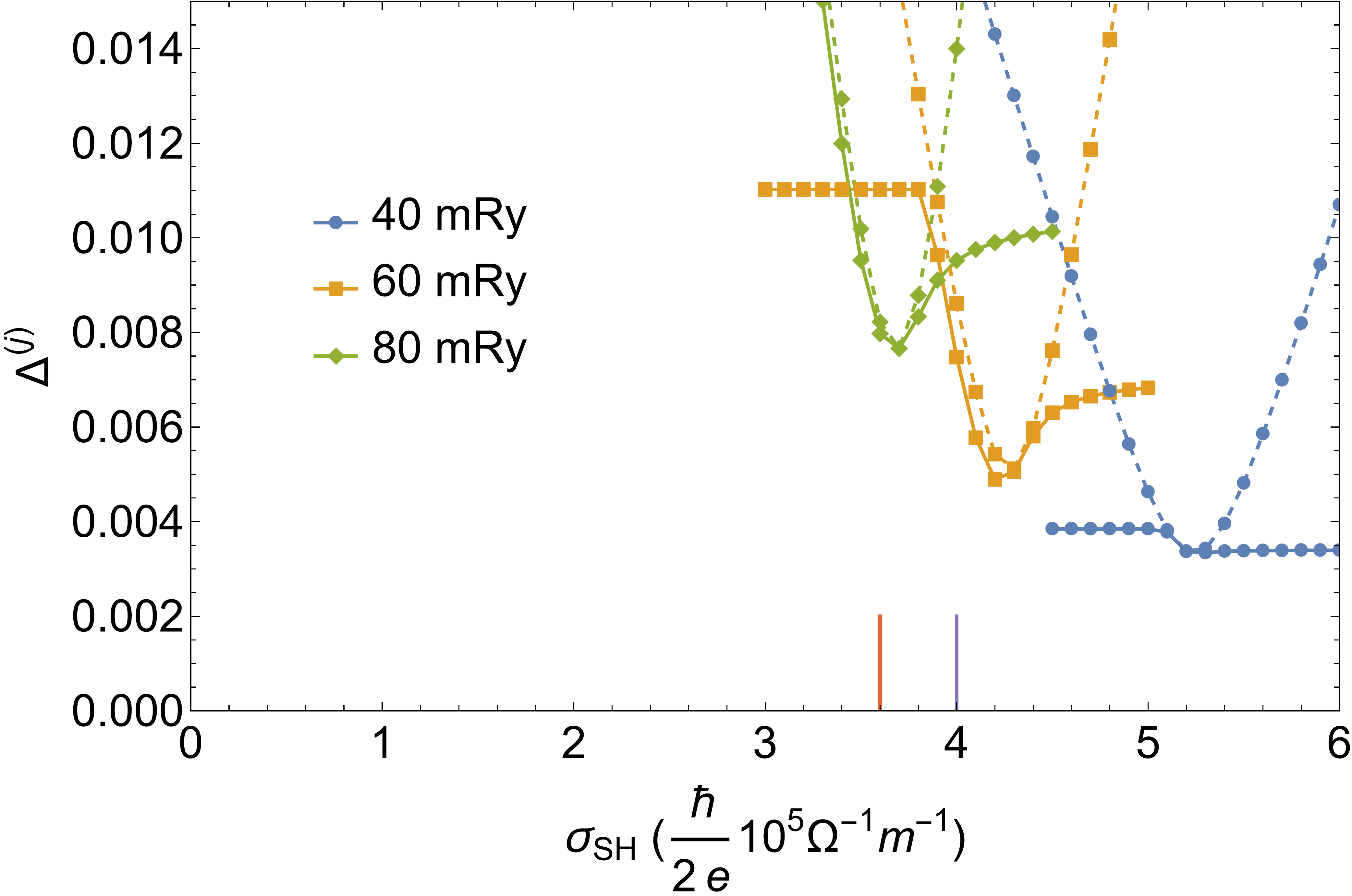}
    \caption{The fitting error for the spin current $\Delta^{(j)}$, see Eq.~(\ref{eq:fitj}), as a function of the bulk spin Hall conductivity at $E_{F}$ for three different values of disorder. The solid curves are optimized over both $l_\mathrm{sf}$ and $\theta_\mathrm{eff}$ and the dotted curves over $\theta_\mathrm{eff}$ with $l_\mathrm{sf}$ set to the optimum value found from the fits to the spin accumulations as in Fig.~\ref{fig:ContourPlots} reported in Table~\ref{tab:coefs}. The purple and red lines give the calculated values for the intrinsic spin Hall conductivity from Ref.~\cite{Guo2008} and Ref.~\cite{salemi2022first} respectively.}
    \label{fig:SpinCurrentFit}
\end{figure}

Figure~\ref{fig:SpinCurrentFit} gives the fits to the spin Hall conductivity shown in Fig.~\ref{fig:jsypt100}(a) using Eq.~(\ref{jssol}) as a function of the bulk spin Hall conductivity $\sigma_\mathrm{SH}$. The fits were done two ways. The solid lines give the fits where the transverse spin-diffusion length $l_\mathrm{eff}^{T}$ and the amplitudes of the exponential variations $\theta_\mathrm{eff}^{(j)}$ were both optimized as a function of the bulk spin Hall conductivity. The dotted curves give the fits in which the transverse spin-diffusion length is fixed to the optimal value found from the fits in Fig.~\ref{fig:ContourPlots}. The first set of fits becomes insensitive to the spin Hall conductivity when its values become small because the transverse spin-diffusion length diverges yielding a position-independent value for the nominally interfacial contribution which then compensates for the variation in the bulk spin Hall conductivity giving a constant fit value. Those fits also continuously lose sensitivity to the spin Hall conductivity when it becomes large because the interfacial contributions overlap giving a roughly parabolic fitting curve that only slowly varies as the bulk spin Hall conductivity changes. Nevertheless, given the agreement between the two fits where both are sensitive, we can conclude that the exponentially varying changes in both the spin accumulation and the spin conductivity are characterized by the same transverse spin-diffusion length.

Figure~\ref{fig:SpinCurrentFit} shows that the bulk spin Hall conductivity extracted from the fits is relatively insensitive to the amount of disorder.  Skew scattering contributions to the spin Hall effect vary as the inverse of the resistivity. Here, for the two large values of disorder, the ratio between the optimum spin Hall conductivities is only 1.16 even though the ratio of the resistivities is 1.42. Also shown are the values of the intrinsic bulk spin Hall conductivities found in two independent calculations.  The insensitivity to disorder and the agreement between the optimal fit values and the independently calculated values indicate that in the interior of the films our calculations capture the intrinsic spin Hall effect despite the differences in the techniques used in the calculations.

\begin{figure}
    \centering
    \includegraphics[width=0.85\columnwidth]{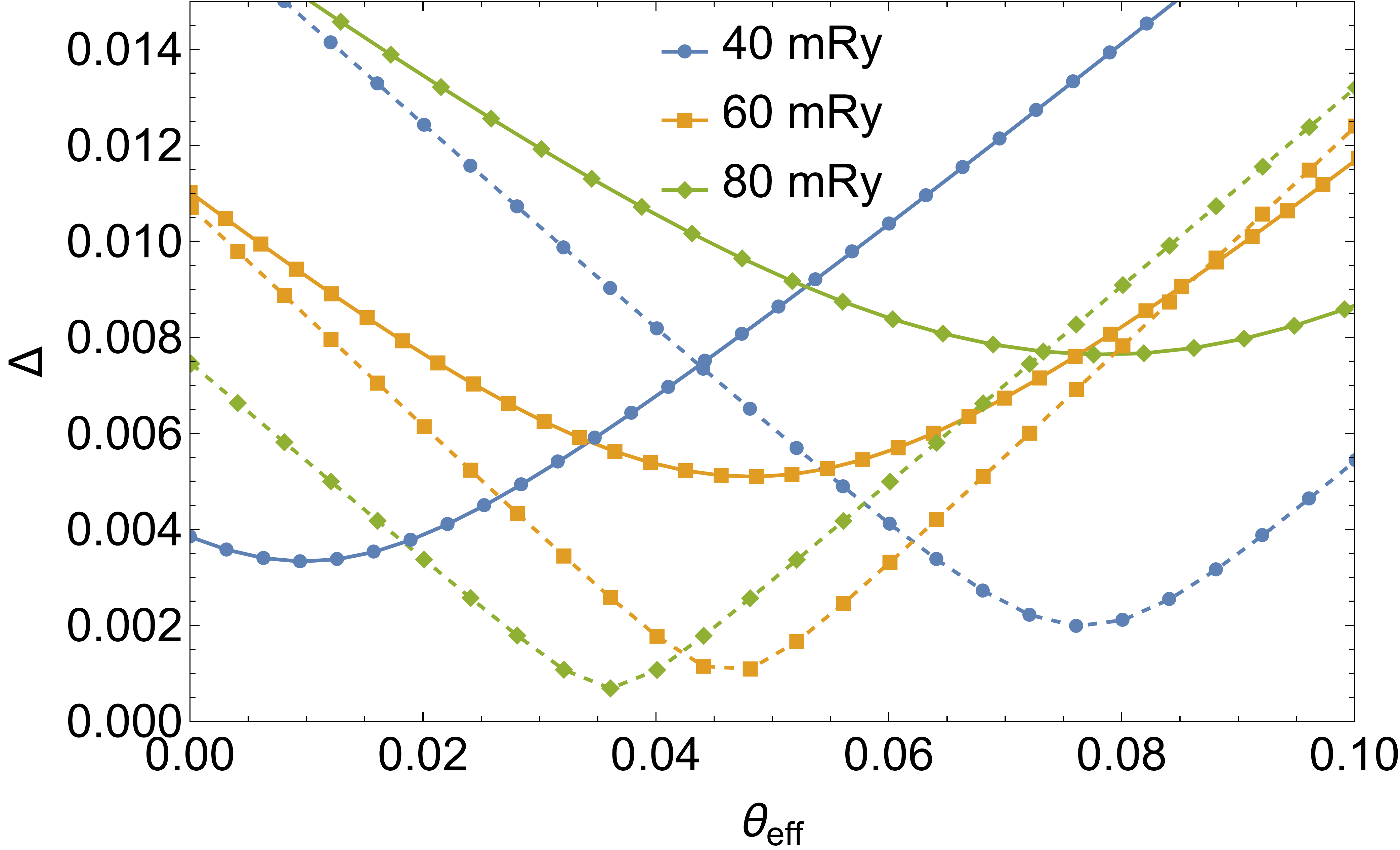}
    \caption{Fitting error for the spin current $\Delta^{(j)}$ (solid lines)  and spin accumulation $\Delta^{(\mu)}$(dashed lines), as a function of $\theta_\mathrm{eff}$ at $E_{F}$ for three different values of disorder. All calculations done at fixed spin-diffusion length as reported in Table~\ref{tab:coefs}. The fits for the spin currernt were optimized over the bulk spin Hall conductivity.}
    \label{fig:Thetaeff}
\end{figure}

While the transverse spin-diffusion lengths are the same for both quantities, their amplitudes are not. Fig.~\ref{fig:Thetaeff} shows the fits to the data in Fig.~\ref{fig:pt100sy}(a) and Fig.~\ref{fig:jsypt100}(a) as a function of the amplitudes of the exponential variations near the interfaces. Both sets of fits use fixed values of the transverse spin-diffusion length and the fits of the spin current data were optimized over the bulk spin Hall conductivity. This figure clearly shows that not only do the amplitudes found from the spin accumulation and the spin Hall conductivity not agree with each other, they have opposite trends as a function of disorder. These results indicate a clear breakdown of the ability of the spin-diffusion model described in Sec.~\ref{sec:SpinDiffusion} to explain our calculated results.

Similar results hold for the fits done at $E'$ except that, because the spin-diffusion lengths are longer, it is more difficult to fit the spin current data meaningfully if both the spin-diffusion length and the bulk spin Hall conductivity are simultaneously used as fitting parameters.

\end{document}